\documentclass[onefignum,onetabnum]{siamart251216}

\usepackage{amsfonts}
\usepackage{graphicx}
\usepackage{epstopdf}
\usepackage{algorithmic}
\usepackage{threeparttable}
\usepackage{booktabs}
\usepackage{multirow}
\usepackage{pdflscape}
\usepackage{amsmath}
\usepackage{braket}
\usepackage{soul}
\usepackage{enumitem}
\usepackage{microtype}

\usepackage{comment}
\definecolor{note}{RGB}{48,96,192}

\usepackage{hyperref}
\hypersetup{
    colorlinks,
    citecolor=blue,
    filecolor=blue,
    linkcolor=blue,
    urlcolor=blue
}
\usepackage{tikz}
\usetikzlibrary{trees}
\usetikzlibrary{shapes}
\usetikzlibrary{arrows.meta}
\usetikzlibrary{positioning}
\usetikzlibrary{calc}

\ifpdf
  \DeclareGraphicsExtensions{.eps,.pdf,.png,.jpg}
\else
  \DeclareGraphicsExtensions{.eps}
\fi

\newcommand{\tableref}[1]{Table~\ref{#1}}
\newcommand{\sectionref}[1]{Section~\ref{#1}}
\newcommand{\tout}{\text{out}}
\newcommand{\tin}{\text{in}}
\DeclareMathOperator{\tr}{tr}
\renewcommand{\P}{ \mathbb P }
\newcommand{\EXP}{\mathbb{E}}
\newcommand{\Var}{\mathbb{V}\mathrm{ar}}
\newcommand{\ud}[1]{\mathrm{d}#1}
\newcommand{\deriv}[3][]{\frac{\ud^{#1} \hspace{-0.3mm} #2}{\ud{#3}^{#1}}}

\newcommand{\Hcal}{\mathcal{H}}
\newcommand{\pnz}{\mathcal{P}} 
\newcommand{\qnz}{\mathcal{Q}} 
\newcommand{\rc}{\rho_C} 
\newcommand{\re}{\rho_E} 
\newcommand{\rt}{\rho} 

\newcommand{\fitTable}[1]{\resizebox{\linewidth}{!}{#1}}

\headers{UQ for QC}{Bennink, Burkovska, Pieper, Ramirez, Wong}

\title{Uncertainty Quantification for \\ Quantum Computing\thanks{Submitted for review.
\funding{This manuscript has been authored by UT-Battelle LLC under contract DE-AC05-00OR22725 with the US Department of Energy (DOE). The US government retains and the publisher, by accepting the article for publication, acknowledges that the US government retains a nonexclusive, paid-up, irrevocable, worldwide license to publish or reproduce the published form of this manuscript, or allow others to do so, for US government purposes. DOE will provide public access to these results of federally sponsored
research in accordance with the DOE Public Access Plan (https://www.energy.gov/doe-public-access-plan).}}}

\author{Ryan Bennink\thanks{Quantum Information Sciences Section, Oak Ridge National Laboratory, (\email{benninkrs@ornl.gov}).}
\and Olena Burkovska\thanks{Mathematics in Computation Section, Oak Ridge National Laboratory (\email{burkovskao@ornl.gov}, \email{pieperk@ornl.gov}).}
\and Konstantin Pieper\footnotemark[3]
\and Jorge Ramirez\thanks{Universidad Nacional de Colombia, Sede Medellin, Colombia (\email{jmramirez@unal.edu.co}).}
\and Elaine Wong\thanks{Corresponding author. Oak Ridge National Laboratory (\email{wongey@ornl.gov}).}
}

\usepackage{amsopn}

\begin{document}

\maketitle

\begin{abstract}
This review is designed to introduce mathematicians and computational scientists to quantum computing (QC) through the lens of uncertainty quantification (UQ) by presenting a mathematically rigorous and accessible narrative for understanding how noise and intrinsic randomness shape quantum computational outcomes in the language of mathematics. By grounding quantum computation in statistical inference, we highlight how mathematical tools—probabilistic modeling, stochastic analysis, Bayesian inference, and sensitivity analysis—can directly address error propagation and reliability challenges in today’s quantum devices. We also connect these methods to key scientific priorities in the field, including scalable uncertainty-aware algorithms and characterization of correlated errors. The purpose is to narrow the conceptual divide between applied mathematics, scientific computing and quantum information sciences, demonstrating how mathematically rooted UQ methodologies can guide validation, error mitigation, and principled algorithm design for emerging quantum technologies, in order to address challenges and opportunities present in modern-day quantum high performance and fault-tolerant quantum computing paradigms.
\end{abstract}

\begin{keywords}
uncertainty quantification, quantum computing, survey
\end{keywords}

\begin{MSCcodes}
81-01, 81-02, 81-08, 68Q09, 81P68, 81P16, 81P18, 81S22, 82C10 
\end{MSCcodes}

\section{Motivation and Introduction}
\label{sec:intro}

Mathematical methods for UQ in QC have yet to be truly applied in a systematic way. This may be ascribed to a conceptual mismatch between fields that do not frequently interact, as well as the rapid evolution of quantum devices based on diverse technologies.
Mathematical abstraction has been proven to be a rigorous way to bridge such gaps. As such, it can provide language to illuminate how noise in presently available Noisy Intermediate-Scale Quantum (NISQ) computers propagates through a computation and affects measured outputs.
This is because such computations are prob\-abilistic in nature and, on currently available NISQ hardware, heavily influenced by undesired systematic and random errors.
In order to interpret the results of every computation, error mitigation strategies or on-device quantum error correcting codes are used, many of which are still at the research stage. This presents a unique opportunity to align mathematical research with the basic needs of the Advanced Scientific Computing Research (ASCR) program under the U.S. Department of Energy (DOE).
For example, to further the development of uncertainty-aware hybrid modeling approaches, there is a need for scalable algorithms that can more efficiently estimate cross-correlations-particularly in high-dimensional settings where naïve methods become computationally prohibitive—in order to provide more accurate and tractable representations for correlation-aware solvers, as well as support for higher-dimensional distributions and adaptive measurement strategies.

UQ is a mathematical discipline focused on the rigorous characterization, propagation, and reduction of uncertainties in mathematical models and computational predictions. It provides a framework to assess the reliability of simulations, particularly when models are driven by incomplete knowledge, stochastic inputs, or noisy data. UQ integrates tools from probability theory, statistics, numerical analysis, and optimization to quantify how uncertainty in inputs leads to variability in outputs. In contemporary science and engineering, UQ is starting to be recognized as a possible enabler of decision-making under uncertainty, as a way to predict credibility of complex simulations---ranging from climate modeling and aerospace design to biomedical imaging and quantum computing---and guiding experimental design and model calibration by identifying the most influential sources of uncertainty. Some examples are highlighted in \tableref{tab:uq_domains}.

\begin{table}[htbp]
\centering
\small
\renewcommand{\arraystretch}{1.4}
\fitTable{%
\begin{tabular}{|p{1.7cm}|p{6cm}|p{5cm}|}
\hline
\textbf{Domain} & \textbf{Role of UQ in Decision Making} & \textbf{Example / Reference} \\
\hline
Earth\quad Systems Modeling & 
Quantifies the spread of predictions from different models and initial conditions, helping policymakers assess the risks of various climate scenarios. & 
Assessing future sea level rise projections to inform coastal infrastructure planning~\cite{ipcc2021}. \\
\hline
Aerospace Design & 
Enables efficient design by accounting for uncertainties in aerodynamics, material properties, and operating environments. Used in reliability-based optimization of structures and flight trajectories. & 
Design of reusable launch vehicles under uncertain aerodynamic loads~\cite{oberkampf2002}. \\
\hline
Biomedical Imaging & 
Reduces diagnostic errors and enables better treatment decisions by quantifying noise and model-based uncertainty in image reconstruction (e.g., computed tomography, magnetic resonance imaging) & 
Bayesian UQ for positron emission tomography (PET) image reconstruction improves tumor detection confidence~\cite{kaipio2005}. \\
\hline
\end{tabular}
}
\caption{The role of uncertainty quantification in decision making across scientific domains.}
\label{tab:uq_domains}
\end{table}

QC is a paradigm of computation that harnesses the principles of quantum mechanics---such as superposition, entanglement, and interference---to process information in fundamentally different ways from \emph{classical} (i.e., conventional digital) computers.
Instead of classical bits, which take values 0 or 1, quantum computers use qubits, which can exist in linear combinations of distinct physical ``states'' that represent the values 0 and 1. This principle extends to any number of qubits, allowing quantum systems to explore a vast space of computational possibilities in parallel. Quantum algorithms leverage interference patterns to amplify correct outcomes and suppress incorrect ones. The formalism of quantum mechanics dictates that, even for perfectly isolated systems, measurement outcomes follow probability distributions, making exact outcomes unknowable in advance. Quantum algorithms therefore produce results that are only accessible through repeated measurements, yielding empirical frequencies that approximate the true underlying probabilities. However, quantum computers are extremely sensitive machines, and even tiny disturbances from their surroundings can disrupt their states.
These disturbances, known as quantum noise, are one of the main obstacles to building reliable quantum technologies. Some noise comes from small imperfections in how quantum operations are realized in the physical hardware (called coherent noise), while other noise comes from random, uncontrollable interactions of the quantum hardware with the environment (incoherent noise).
Both types can cause errors in storing, processing, or measuring quantum information. Understanding and controlling quantum noise is essential because it directly impacts how well quantum computers can scale, and it motivates the development of better error management strategies to protect information from being lost. 

From the UQ standpoint, QC is unusual among other modes of computation because the output is always a probability distribution over possible outcomes that must be estimated from data. This is the case even in the absence of model error, numerical error, or experimental noise, since the output is not a deterministic quantity to be computed once, but a random event determined by the internal, unobservable state of the quantum hardware.
The estimation target is not a single number but a statistical object---a distribution or an expectation value---and thus every quantum computation is, at its core, a statistical inference problem: it is \textit{inherently} about quantifying, controlling, and interpreting uncertainty---both the intrinsic randomness of quantum measurement and the statistical error from finite sampling---placing it naturally within the scope of modern UQ theory.
To drive this point home, whereas classical computational science uses UQ to assess and reduce uncertainties that are typically extrinsic (model approximations, floating-point error, parameter uncertainty), QC has intrinsic, irreducible randomness that makes UQ methods not just auxiliary but central to interpreting results at all levels of the computational workflow.

In the preparation of this report, we put forth the premise that classical simulators and UQ algorithms (ranging from forward UQ in the form of sampling to inverse UQ in the form of calibration of error models) will be essential in the study of scalable quantum computing.
\textbf{Our goal} is to narrow the conceptual divide between applied mathematics and quantum information science by providing an instructive, rigorous, and self-contained introduction to quantum computing tailored to mathematicians, along with a representative (though not exhaustive) overview of relevant mathematical methods, including probabilistic modeling, stochastic analysis, Bayesian inference, and sensitivity analysis.
Towards this goal, we view QC through a ``UQ-lens'', which allows to investigate how the state-of-the-art in UQ can address error related challenges and opportunities presented by QC.
Moreover, we are of the opinion that quantum computation can be viewed fundamentally as an important application problem in uncertainty quantification, and we hope to convince the reader that this view will open up possibilities as the field of quantum computing progressesfrom NISQ devices to fault-tolerant (noise-controlled) computation.
In doing so, we underscore the potential for UQ methods not only to diagnose limitations and guide error mitigation in near-term quantum devices but also to enable more principled approaches to quantum algorithm design, validation, and certification.

In Section~\ref{sec:overview}, we give a mathematical overview of QC, introducing notation and basic vocabulary that is used throughout.
In Section~\ref{sec:UQmethods}, we present a selected list of UQ methods currently used in QC in practice.
This list is based on our interests and we do not claim that it includes every aspect of this topic, but rather can be used as an inspirational starting point.
We end with Section~\ref{sec:opportunities}, where we present opportunities and challenges in QC that modern UQ can address.


\section{Overview of Quantum Computing}
\label{sec:overview}

Here we present a self-contained introduction to quantum computing. Our emphasis is the mathematical description of the sources and models for uncertainty in gate-based quantum computation. This is standard textbook material. We follow mostly~\cite{nielsen2010quantum}.

\subsection{Pure states, quantum bases and operators}

Gate-based quantum computing formalizes computation in terms of the controlled evolution of quantum states in a finite-dimensional complex Hilbert space. For $n$ qubits, the state space is $\mathcal{H}_n = (\mathbb{C}^2)^{\otimes n}$, which is a $2^n$-dimensional Hilbert space of complex vectors with complex coefficients. The canonical basis of this space is given by the following $2^n$ states, which are called the \textit{computational basis}: 
\begin{equation}
    \mathcal{B}_n = \{\,\ket{x}: x \in \{0,1\}^n\,\} =\{\,\ket{00\cdots 0}, \ket{00\cdots 01}, \cdots \ket{11\cdots 1}\,\}.
\end{equation}
There are two equivalent ways of mathematically representing each state in $\mathcal{B}_n$. One is as canonical vectors in $\mathbb{C}^{2^n}$. For example, for $n=3$, the state $\ket{101}$ is the sixth vector of $\mathcal{B}_3$ in lexicographic order, and corresponds to the canonical vector $e_6 = [0,0,0,0,0,1,0,0] \in \mathbb{C}^8$. The second one, more relevant to quantum computation, is as vectors in the tensor product $(\mathbb{C}^2)^{\otimes n}$. Specifically, using  the single-qubit vectors $\ket{0} = [1,0]$, $\ket{1}=[0,1]$ in $\mathbb{C}^2$, we write:
\begin{equation}\label{eq:example_tensor}
    \ket{101} = \ket{1} \otimes \ket{0} \otimes \ket{1} = [0,1] \otimes [1,0] \otimes [0,1] \in (\mathbb{C}^2)^{\otimes 3}.
\end{equation}

A \textit{pure quantum state} is a unitary linear combination (or \textit{superposition}) of the computational basis states with complex coefficients whose squared amplitudes add to one:
\begin{equation}\label{eq:def_purestate}
    \ket{\psi} = \sum_{x \in \{0,1\}^n} \alpha_x \ket{x}, \quad \sum_{x \in \{0,1\}^n} |\alpha_x|^2 = 1.
\end{equation}

The \textit{bracket notation} is a slick computational device to perform algebraic computations in $\mathcal{H}_n$. A pure state written as the ``ket'' $\ket{\psi}$ can be understood as a column vector, and the corresponding ``bra'' $\bra{\psi}$ as its transpose conjugate.
Consequently, the inner product between states $\ket{\psi},\ket{\phi}$ is written with the ``bracket'' $\braket{\psi|\phi} \in \mathbb{C}$. It follows from Eq.~\eqref{eq:def_purestate} that $\braket{\psi|\psi} = 1$ for all pure states. The ``ketbra'' between states also plays an important role, as we will see when we define mixed states in \sectionref{sec:mixed_states}. It is written as $\ket{\psi}\bra{\phi} \in \mathbb{C}^{2^n\times 2^n}$ and represents a rank-1 matrix. 

Quantum computations are usually set up with respect to the computational basis but measured with respect to a variety of quantum bases. In general, a \textit{quantum basis} for $\mathcal{H}_n$ is an orthonormal basis for $\mathbb{C}^{2^n}$, namely a collection of pure states $\mathcal{B} = \{\ket{\phi_i}\}_{i=1}^{2^n}$ such that $\braket{\phi_i| \phi_j} = \delta_{i,j}$. As in any other Hilbert space, expressing the arbitrary state $\ket{\psi}$ of Eq.~\eqref{eq:def_purestate} with respect to the basis $\mathcal{B}$ means computing the inner products $\beta_i = \braket{\phi_i| \psi}$ and writing $\ket{\psi} = \sum_{i=1}^{2^n} \beta_i \ket{\phi_i}$. Because states are unitary, we must have $\sum_i |\beta_i|^2 =1$. The $2^n$ complex coefficients $\{\beta_i\}$ are called the \textit{probability amplitudes} of $\ket{\psi}$ with respect to the basis $\mathcal{B}$.

The modulus-squared amplitudes of a state can be interpreted as ``probabilities'' quantifying how the state is distributed across the $2^n$ configurations in the basis. Note however, that they depend on the basis, and are thus not a probability distribution intrinsic to the state.  For example, the 2-qubit ``Bell state'' is defined in terms of the computational basis $\mathcal{B}_2$ as
\begin{equation}\label{eq:Bell}
    \ket{\Phi^+} = \frac{1}{\sqrt{2}} \ket{00} + \frac{1}{\sqrt{2}} \ket{11}.
\end{equation}
We say that $\ket{\Phi^+}$ is in \textit{balanced superposition} because it assigns  probabilities equal to $1/2$ to two elements of $\mathcal{B}_2$. However, $\ket{\Phi^+}$  is part of another important basis in $\mathcal{H}_2$, the ``Bell basis'':
\begin{equation}
    \mathcal{B}^{\text{Bell}} = \left\{ 
        \frac{1}{\sqrt{2}}\left(\ket{00} + \ket{11}\right),
        \frac{1}{\sqrt{2}}\left(\ket{00} - \ket{11}\right),
        \frac{1}{\sqrt{2}}\left(\ket{01} + \ket{10}\right),
        \frac{1}{\sqrt{2}}\left(\ket{01} - \ket{10}\right)
    \right\}.
\end{equation}
With respect to this basis, $\ket{\Phi^+}$ has only one non-zero amplitude, and is not in balanced superposition.

\subsubsection{Tensor products and entanglement}

When working with $n>1$ qubits, we saw in Eq.~\eqref{eq:example_tensor} that there is an under\-lying $n$-fold tensor product operation. Given any $n$ single-qubit states $\ket{\psi_i} = \alpha^{(i)}_0 \ket{0} + \alpha^{(i)}_1 \ket{1}$, for $i=1,\dots,n$, we can form a state in $\mathcal{H}_n$ via $\ket{\psi} = \ket{\psi_1} \otimes \dots \otimes \ket{\psi_n} $. The $2^n$ amplitudes of $\ket{\psi}$ with respect to the computational basis $\mathcal{B}_n$ are indexed by $x \in \{0,1\}^n$ and can be computed as $\alpha_x = \alpha^{(1)}_{x_1} \dots \alpha^{(n)}_{x_n}$. For example, consider the tensor product between the the Bell state in Eq.~\eqref{eq:Bell} and the ``Minus'' state $\ket{-} = \frac{1}{\sqrt{2}}(\ket{0} - \ket{1}) \in \mathcal{H}_1$ . We get the following state in $\mathcal{H}_3$,
\begin{equation}\label{eq:Phi+-}
    \ket{\Phi^+}\otimes \ket{-}= \frac{1}{2} \ket{000} - \frac{1}{2}\ket{001} + \frac{1}{2}\ket{110} - \frac{1}{2}\ket{111}.
\end{equation}

A defining feature of quantum computing is that $\mathcal{H}_n$ is not simply the collection of all $n$-way tensor products of single-qubit states. The states in $\mathcal{H}_n$ that can be written as tensor products,
\begin{equation}
    \ket{\psi} = \ket{\psi_1} \otimes \dots \otimes \ket{\psi_n} 
\end{equation}
where each $\ket{\psi_i}$ is a pure state in $\mathcal{H}_1$, are called \textit{separable}. But for example, the Bell state in Eq.~\eqref{eq:Bell} cannot be written as a tensor product of any pair of single-qubit states. This phenomenon is called \textit{quantum entanglement}, and we say that $\ket{\Phi^+}$ entangles its two qubits. We will see that when measuring $\ket{\Phi^+}$, knowing the outcome of one of its qubits immediately constrains the outcome of the other. Entangled states exhibit correlations that are an essential quantum resource enabling exponential state space, quantum parallelism, and the possibility of error-correcting codes beyond classical limits. 

The situation in Eq.~\eqref{eq:Phi+-} is common in quantum circuits and helps us introduce the notion of \textit{subsystems} and \textit{partial entanglement}. The state $\ket{\psi} = \ket{\Phi^+}\otimes \ket{-}$ is entangled in $\mathcal{H}_3$ because it is not the tensor product of three pure states. However, we could think of $\mathcal{H}_3$ as partitioned between two subsystems $\mathcal{H}_3 = \mathcal{H}_2 \otimes \mathcal{H}_1$ and $\ket{\psi}$ is separable across this partition because is a product of two pure states, one from each subsystem. It is thus partially entangled, and its entanglement is localized inside subsystem $\mathcal{H}_2$. 

\textit{Schmidt's decomposition} theorem is crucial in this context.
Suppose $\mathcal{H}_n$ is partitioned into two subsystems $\mathcal{H}_m$ and $\mathcal{H}_{n-m}$ for some $m<n$.
Then, there exist $r \geq 1$, real amplitudes $\{\alpha_i\}_{i=1}^r$ and orthonormal sets $\{\ket{\phi_i}\}_{i=1}^r \subset \mathcal{H}_m$, $\{\ket{\varphi_i}\}_{i=1}^r \subset \mathcal{H}_{n-m}$ such that
\begin{equation}\label{eq:SchmidtDec}
    \ket{\psi} = \sum_{i=1}^r \alpha_i \ket{\phi_i} \otimes \ket{\varphi_i} \quad \text{and} \quad \sum_i |\alpha_i|^2 = 1.
\end{equation}

Equation \eqref{eq:SchmidtDec} is called the Schmidt decomposition $\ket{\psi}$. The number $r$ is called the Schmidt rank, and helps determine the type of entanglement across the partition. For example, across the partition $\mathcal{H}_3 = \mathcal{H}_2 \otimes \mathcal{H}_1$, the state in Eq.~\eqref{eq:Phi+-} has Schmidt decomposition simply as $\ket{\psi} =\ket{\Phi^+}\otimes \ket{-} $, so $r=1$ and we have separability. But across $\mathcal{H}_3 = \mathcal{H}_1 \otimes \mathcal{H}_2$, the state is decomposed as
\begin{equation*}
    \ket{\psi} = \frac{1}{\sqrt{2}} \ket{0} \otimes (\ket{0} \otimes \ket{-}) + \frac{1}{\sqrt{2}} \ket{1} \otimes (\ket{1} \otimes \ket{-}),
\end{equation*}
so $r=2$ and $\ket{\psi}$ is entangled.

\subsubsection{Quantum operators}

Quantum computing is defined by the action of linear operators over states in $\mathcal{H}_n$. A \textit{quantum operator} is a linear map $A: \mathcal{H}_n \to \mathcal{H}_n$, and the action of it on a pure state $\ket{\psi}$ is denoted as $A \ket{\psi} = \ket{A \psi}$. Given any quantum basis $\mathcal{B} = \{\ket{\phi_i}\}_{i=1}^{2^n}$ for $\mathcal{H}_n$, the set of rank-one operators  $\{\ket{\phi_i}\bra{\phi_j}\}_{i,j=1}^{2^n}$ is a basis for the space of operators. Then any operator $A$ can be represented as matrices in $\mathbb{C}^{2^n \times 2^n}$ with respect to any quantum basis and written in the convenient operator expansion formula
\begin{equation}\label{eq:operatorExpansion}
    A = \sum_{i,j=1}^{2^n} A_{i,j} \ket{\phi_i} \bra{\phi_j}, \quad A_{i,j} = \braket{\phi_i| A | \phi_j}.
\end{equation}

\tableref{tab:operators} summarizes the most important classes of operators in quantum computing, which can be subdivided into two broad classes depending on the relation between the operator $A$ and its conjugate-transpose $A^\dag$: unitary or Hermitian.
Hermitian operators, defined by $A^\dagger = A$, have real eigenvalues, which in quantum computing correspond to physically meaningful observations.
Unitary operators, defined by $A^\dagger = A^{-1}$, are norm-preserving and are used for evolving quantum systems.
This evolution can be a discrete operation defined by a unitary matrix $U$, a \textit{gate} in a quantum circuit, or can depend on a continuous-time $t>0$ realized by $U(t) = e^{-i H t}$ for some Hermitian Hamiltonian operator $H$.
Unitary operators also preserve orthonormality, so any quantum basis can be obtained as a unitary transformation of another basis.

As is the case for states, operators in $\mathcal{H}_n$ that can be constructed as a tensor product of subsystems are called \textit{separable operators}. But the most interesting operators in multi-qubit quantum computing, Hermitian or unitary, are not separable. A canonical example in $\mathcal{H}_2$ is the CNOT gate, which can be written in the computational basis as
\begin{equation}\label{eq:CNOT}
    U_{\text{CNOT}} = \left[\begin{array}{cccc}
        1 & 0 & 0 & 0  \\
        0 & 1 & 0 & 0\\
        0 & 0 & 0 & 1\\
        0 & 0 & 1 & 0\\
    \end{array}\right]
\end{equation}
This gate flips the second qubit when the first qubit is equal to 1. For example $U \ket{01} = \ket{01}$ but $U \ket{11} = \ket{10}$. It, however, cannot be written as a tensor product of any pair of unitaries in $\mathcal{H}_1$.

\begin{table}[h!]
\centering
\small
\renewcommand{\arraystretch}{1.3}
\fitTable{%
\begin{tabular}{|l|c|p{5.5cm}|}
\hline
\textbf{Operator type} & \textbf{Defining property} & \textbf{Physical / computational role} \\
\hline
\textbf{Unitary} $U$ 
& $U^\dagger U = U U^\dagger = I$ 
& Describes reversible \textit{quantum evolution} and \textit{quantum gates}. \\

\hline
\textbf{Observable} $O$ 
& $O^\dagger = O$ 
& Represents a \textit{measurable quantity}. Has real eigenvalues and orthogonal eigenvectors; the eigenvalues correspond to possible measurement outcomes. \\

\hline
\textbf{Positive} $M$ 
& $\braket{\psi|M|\psi} \geq 0$ for all $\ket{\psi} \in \mathcal{H}_n$
& Special case of observable. Represents a possible measurement outcome in a POVM.  Also called effect operator.  \\

\hline
\textbf{Projection} $P$ 
& $P^2 = P = P^\dagger$ 
& Special case of positive operator. Describes an \textit{ideal measurement outcome} or \textit{event}. 
Form Projective Measurements (PVMs). \\

\hline
\textbf{Density} $\rho$ 
& $\rho = \rho^\dagger,\; \rho \ge 0,\; \mathrm{Tr}(\rho) = 1$ 
& Represents a general (possibly mixed) \textit{quantum state}. 
Encodes statistical mixtures and allows computation of expectations via $\langle A \rangle = \mathrm{Tr}(A \rho)$. \\

\hline
\end{tabular}
}
\caption{Main types of operators in quantum computing: definitions and roles.}
\label{tab:operators}
\end{table}

\subsubsection{Measurements on pure states}
\label{sec:meas_pure}

The operators labeled as Observable, Positive and Projection in \tableref{tab:operators} are all related to the act of \textit{quantum measuring}, which is the process by which information can be extracted from a quantum computation. A more general picture will be given in \sectionref{sec:mixed_states}, but for now we can illustrate the case of projection operators onto single qubits, which correspond to the simplest case of observables and are very common operations in quantum circuits. 

Consider the pure state $\ket{\psi} = \alpha_0 \ket{0} + \alpha_1 \ket{1} \in \mathcal{H}_1$ and suppose we want to ``measure it with respect to the computational basis'' $\mathcal{B}_1$.
Physically, that means that we will make the state interact with a measurement device that can determine whether $\ket{\psi}$ is in the $\ket{0}$ or $\ket{1}$ state. These are the two possible outcomes of the measurement, and they occur with with probabilities $|\alpha_0|^2$ and $|\alpha_1|^2$, respectively. These probabilities can be recovered from $\ket{\psi}$ using the \textit{projection operators} along each member of the computational basis: $P_0 = \ket{0}\bra{0}$ and $P_1 = \ket{1}\bra{1}$ which are exhaustive because they satisfy $P_0+P_1 = I$. We write
\begin{equation}\label{eq:probMx}
    \P_{\ket{\psi}}(\text{observe } 0) = \braket{\psi|P_0|\psi} = \braket{\psi|0}\braket{0|\psi} = |\braket{0|\psi}|^2 = |\alpha_0|^2.
\end{equation}
And similarly for $\P_{\ket{\psi}}(\text{observe } 1)$. Expression Eq.~\eqref{eq:probMx} is the first instance where we have used the symbol $\P$ for a probability: it denotes the \textit{quantum probability} that our device will observe that $\ket{\psi}$ is in state $\ket{0}$ upon measuring. We will say more about quantum probabilities and their relation to classical probabilities in \sectionref{sec:quantum_probs}.

A crucial difference between quantum and classical systems lies in the nature of measurement: observing a quantum state fundamentally alters it. When $\ket{\psi}$ is measured in the computational basis, its superposition \textit{collapses} into one of the definite states $\ket{0}$ or $\ket{1}$, and all the quantum uncertainty encoded in its amplitudes disappears. This property is central to the probabilistic interpretation of quantum mechanics—it is what allows quantum circuits to yield classical outcomes that can be read out and statistically analyzed. However, it also introduces a major source of error in quantum computation: because measurement irreversibly destroys coherence, any unintended interaction with the environment that acts as an implicit measurement can collapse the state prematurely, leading to \textit{decoherence} and loss of quantum information.
  
Single qubit projective measurements can be performed on separate components of a multi-qubit system. For example, suppose we want to measure the first qubit of $\ket{\psi} \in \mathcal{H}_n$ with respect to the computational basis. Expand $\ket{\psi}$ as
\begin{equation}
\label{eq:state_before_measurent}
    \ket{\psi} = \ket{0}\otimes \ket{\psi_0} + \ket{1}\otimes \ket{\psi_1}
\end{equation}
where $\ket{\psi_i} = (\bra{i} \otimes I) \ket{\psi} \in \mathcal{H}_{n-1}$ are state vectors in the Hilbert space of the remaining $n-1$ qubits. Applying the measurement $\{P_0,P_1\}$, with $P_i = \ket{i}\bra{i} \otimes I$, to the first qubit collapses it to an outcome of $\ket{0}$ or $\ket{1}$ with probabilities $\|\ket{\psi_0}\|^2,\|\ket{\psi_1}\|^2$. That component of the state becomes a \textit{classical register} -- a known bit of information with value $0$ or $1$. The post-measurement state in $\mathcal{H}_n$ is
\begin{equation}
\label{eq:state_after_measurent}
    \ket{0}\otimes \frac{\ket{\psi_0}}{\| \ket{\psi_0} \|} \quad \text{or} \quad \ket{1}\otimes \frac{\ket{\psi_1}}{\| \ket{\psi_1} \|}
\end{equation}
depending on the outcome of the measurement. It combines the value of the first qubit that was observed in the measurement with the remaining un-measured and uncertain $n-1$ qubits. In statistics, the measurement process is the transition of the system from the state $\ket{\psi}$ given in Eq.~\eqref{eq:state_before_measurent} to a random distribution in the state space with the states given in Eq.~\eqref{eq:state_after_measurent}.

\subsection{Quantum circuits and algorithms}

A \textit{quantum algorithm} is a computational procedure designed to run on a quantum computer. It uses quantum circuits as its fundamental building blocks. Modern quantum computers are intricate systems comprising a quantum processing unit, gate-generation and control hardware, measurement and readout circuitry, and mechanisms for quantum error detection, correction, and memory~\cite{gill2022quantum}. 

In the current NISQ era, hardware is limited by short coherence times, restricted qubit connectivity, gate infidelities, and the absence of full quantum error correction. These constraints strongly influence which algorithms can be realistically executed and have motivated the development of shallow, noise-resilient, or hybrid quantum–classical approaches. As a result, only a small number of algorithms have demonstrated potential for practical, industry-relevant advantages. Notable examples include the quantum Fourier transform (QFT), Grover’s search algorithm, variational quantum eigensolvers (VQE), the quantum approximate optimization algorithm (QAOA), and various quantum error mitigation (QEM) techniques. We will refer to  some of these algorithms in this report. For more basic details we refer the reader to~\cite[Chapter~4]{nielsen2010quantum}. For a contemporary review of the state-of-the-art in quantum circuits, their industrial applications an challenges; see~\cite{sood2025industrial,singh2023contemporary}. 

\subsubsection{Quantum circuits}

A \textit{quantum circuit} is a structured way of specifying a unitary operator that performs a desired computational task on a collection of qubits. It describes the idealized time evolution of a set of isolated qubits, where dynamics are governed solely by a sequence of unitary transformations, called \textit{gates}, prescribed in the circuit.  In practice, each gate in a circuit corresponds to a particular basic action on the quantum hardware, where physical processes governing the qubits' dynamical evolution are controlled to (approximately) implement the corresponding unitary operator. Thus, a quantum circuit provides both an abstract computational description and a blueprint for how hardware attempts to implement the intended evolution.

\begin{figure}[htbp]
    \centering
    \includegraphics[width=\linewidth,keepaspectratio]{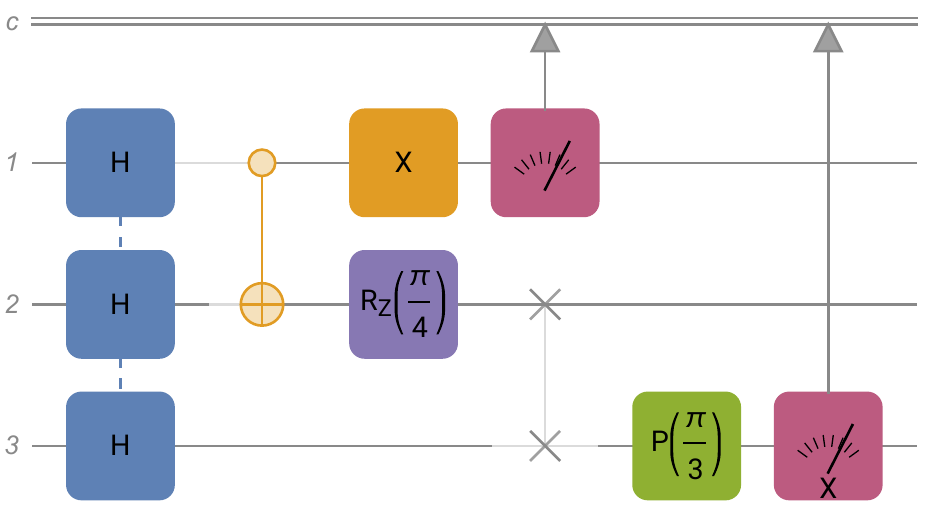}
    \caption{Example of a quantum circuit on three qubits}
    \label{fig:circuit}
\end{figure}

Mathematically, a quantum circuit on $n$ qubits prescribes a sequence of unitary and measurement operators on subsystems of $\mathcal{H}_n$. It is represented graphically as in Figure \ref{fig:circuit} with symbols that are drawn from a library of operators and conventions that are for the most part consistently used across the literature. 

The numbered horizontal lines are called ``wires'' and there is one for each qubit. The circuit is read from the left. The first operator to act is a Hadamard gate on all three qubits, which is given by the operator
\begin{equation}
    U_1 =  \left[\begin{array}{cc}
         \frac{1}{\sqrt{2}}&\frac{1}{\sqrt{2}}  \\
         \frac{1}{\sqrt{2}}&- \frac{1}{\sqrt{2}} 
    \end{array}\right]^{\otimes 3}
\end{equation}
These gates perfectly entangle all three qubits (if they are initialized, e.g., in the $\ket{0}$ state). Then a CNOT gate acts on the first two qubits. That corresponds to operator $U_2= U_{\text{CNOT}} \otimes I$ with $U_{\text{CNOT}}$ as in Eq.~\eqref{eq:CNOT} and the identity. Then we have a separable gate with an X operator on the first qubit, a ``rotation along the $z$-axis by $\pi/4$'' on the second qubit and the identity on the third:
\begin{equation}
    U_3 = U_X \otimes U_{R_z(\pi/4)} \otimes I = 
    \left[
    \begin{array}{cc}
        0 & 1 \\
        1 & 0
    \end{array}
    \right] \otimes 
    \left[
        \begin{array}{cc}
         e^{-\frac{i \pi}{8}} & 0 \\
         0 & e^{\frac{i \pi }{8}} \\
        \end{array}
\right] \otimes 
\left[
    \begin{array}{cc}
        1 & 0 \\
        0 & 1
    \end{array}
    \right].
\end{equation}
A ``SWAP" gate swaps exchanges the positions of qubits two an three. This is represented by the line with two xs, and is given by the operator 
\begin{equation}
   U_4 = I \otimes U_{\text{SWAP}} = I \otimes \left[
\begin{array}{cccc}
 1 & 0 & 0 & 0 \\
 0 & 0 & 1 & 0 \\
 0 & 1 & 0 & 0 \\
 0 & 0 & 0 & 1 \\
\end{array}
\right].
\end{equation}
The $P(\pi/3)$ denotes a ``phase-shift'' by an angle of $\pi/3$ on the third qubit
\begin{equation}
    U_5 = I_2 \otimes \left[
        \begin{array}{cc}
         1 & 0 \\
         0 & e^{\frac{i \pi }{3}} \\
        \end{array}
\right].
\end{equation}

The circuit in \ref{fig:circuit} defines a unitary operator $U$ composed of five \textit{layers} for which the output state can be computed as
\begin{equation}
    \ket{\psi_{\tout}} = U \ket{\psi_{\tin}}= U_5~U_4 ~ U_3 ~ U_2 ~ U_1 \ket{\psi_{\tin}}.
\end{equation}
Note that in $U$, the operators in each layer composed by matrix multiplication and are written in the opposite direction as in the graphical representation.

Finally, the circuit in Figure~\ref{fig:circuit} defines a probability distribution based on the possible values of its measurements. In this case, there are two measurements with two outcomes each, which take place after all the unitaries are applied. The first measurement is of the first qubit with respect to the computational basis as described in \sectionref{sec:meas_pure}. It is not influenced by SWAP and phase-shift of the second and third qubits. The result of that measurement is sent to the double wire on top that represents the \textit{classical register} with the measurement's output. For the second measurement, we ``measure the third qubit with respect to the X basis''. That means applying the projector $\{P_{\ket{+}},P_{\ket{-}}\}$ along the basis formed by the eigenstates of the operator $U_X$. This is also sent to a classical register as a second bit of information.

The recorded values in the classical register can be $\{0,+\}$, $\{0,-\}$, $\{1,+\}$ or $\{1,-\}$. For example, if we apply the circuit gates on $\ket{\psi_{\tin}} = \ket{000}$, we have just before measurement:
\[
\ket{\psi_{\tout}} = \tfrac{1}{\sqrt{8}}\left(z_0(\ket{000}+\ket{010}+\ket{100}+\ket{110})+z_1p(\ket{001}+\ket{011}+\ket{101}+\ket{111})\right),
\]
where $z_0=e^{-i\pi/8},z_1=e^{i\pi/8}$ and $p=e^{i\pi/3}$.
Expressing $\ket{\psi_{\tout}}$ with respect to the measurement bases, we can then calculate the corresponding probabilities of the outcomes:
\begin{align}
    \P_{\ket{\psi_{\tout}}}(\text{observe} \{0,+\}) &= \P_{\ket{\psi_{\tout}}}(\text{observe} \{1,+\} )=\tfrac18 |z_0+z_1p|^2\approx 0.185, \\
    \P_{\ket{\psi_{\tout}}}(\text{observe} \{0,-\}) &= \P_{\ket{\psi_{\tout}}}(\text{observe} \{1,-\})=\tfrac18 |z_0-z_1p|^2 \approx 0.315.
\end{align}

It is important to note that real quantum circuits bear little resemblance to that shown in Figure \ref{fig:circuit}. State-of-the-art quantum algorithms operate on tens to hundreds of qubits and, once decomposed into hardware-native gates, routinely demand circuits with thousands to billions of gate operations and intricate patterns of entangling interactions. However, these large quantum circuits are assembled from a small number of simple circuit blocks (like the ones in our example) chosen to match the structure of the computational task and the constraints of the underlying hardware. 

\subsubsection{Variational quantum circuits}\label{sec:VQC}
Variational quantum circuits---also known as parametrized quantum\linebreak[1] circuits---are families of quantum circuits whose unitary transformation depends smoothly on a vector of real parameters, $U = U(\theta)$ for $\theta \in \mathbb{R}^p$.
The general goal is efficiently evaluate a scalar \textit{cost function}
\begin{equation}\label{eq:defCost}
    C(\theta) = \braket{\psi_0 | U(\theta)^\dag OU(\theta)| \psi_0}
\end{equation}
where $O$ is some observable and $\ket{\psi_0}$ is the initial state. These evaluations are incorporated into a  \textit{hybrid quantum-classical algorithm} in which a classical optimizer provides updates of $\theta$ aimed at minimizing $C(\theta)$.

In this context, the circuit's unitary is called an \textit{ansatz}. It is typically expressed as a product of layers
\begin{equation}
    U(\theta) = U_L(\theta_L) \cdots U_1(\theta_1),\quad
    U_\ell(\theta_\ell) = W_\ell \prod_{j=1}^{m_\ell} e^{-i \theta_{\ell,j} H_{\ell,j}}, \quad \theta_{\ell,j} \in [-\pi,\pi],
\end{equation}
where each $H_{\ell,j}$ is a Hermitian operator and each $W_{\ell}$ is an unparametrized unitary. For a given cost function, the ansatz is chosen to be hardware efficient, so that it can be implemented with a low number of layers on the target device. It is also important that the ansatz is expressive, meaning that $U(\theta)\ket{\psi_0}$ can produce a sufficiently rich family of states as $\theta$ varies.

\begin{figure}[htbp]
    \centering
    \includegraphics[width=\linewidth,keepaspectratio]{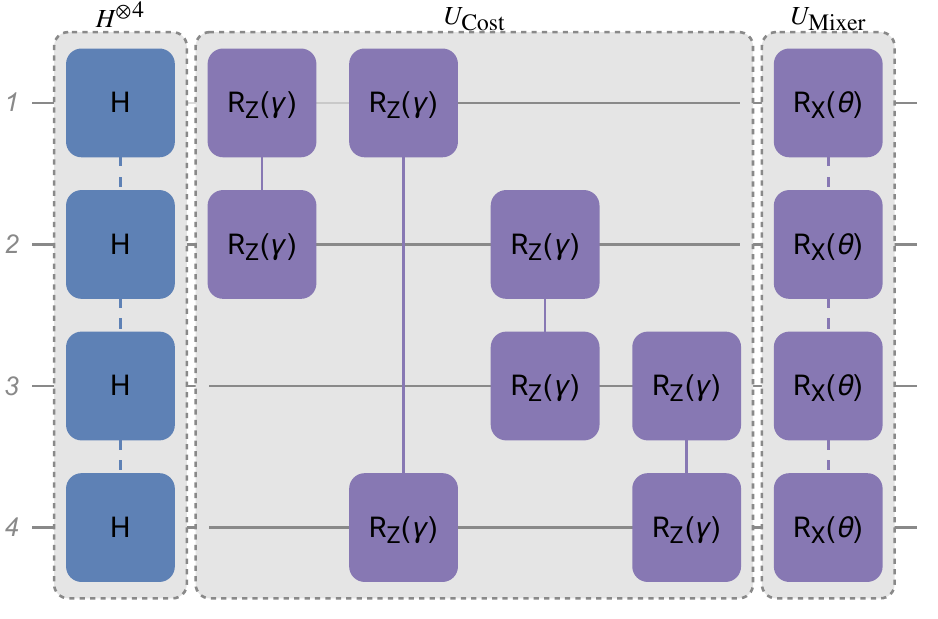}
    \caption{Example of a variational quantum circuit on four qubits.}
    \label{fig:qaoa}
\end{figure}

Figure \ref{fig:qaoa} shows the diagram of a small Quantum Approximate Optimization Algorithm (QAOA) on four qubits. This circuit is used for combinatorial optimization problems. This ansatz in particular, can be used to solve the MaxCut problem in graph theory~\cite{rodriguez2025QAOA}. After an entanglement layer, each of the blocks in the group $U_{\text{Cost}}$ denotes the operator $R_Z(\gamma)=e^{-i \gamma Z \otimes Z}$ acting on specific pairs of qubits. The operators in the mixer layer are $R_X(\theta)=e^{-i \theta X}$. The parameters of this ansatz are $\theta, \gamma$. The observable of the QAOA is a measurement of the each qubit in the computational basis. A realistic industrial QAOA implementation on current NISQ processors generally involves tens of qubits and one to three layers like that shown in Figure \ref{fig:qaoa}. For this algorithms, noise and finite coherence times limit circuit depth~\cite{harrigan2021quantum}.

A central feature of variational quantum circuits is that they allow efficient computation of the gradient of the cost function with respect to $\theta$, enabling the use of gradient-based classical optimizers. This formula is called the \textit{parameter-shift rule} and is described in detail in~\cite{schuld2019evaluating}. It is based on the observation that for a unitary of the form $U(\theta_j) = e^{- i \theta_{j} H_j}$, where $H_j$ has two eigenvalues $\pm r$,  the gradient of the cost function $C(\theta)$ in Eq.~\eqref{eq:defCost} can be computed with shifted evaluations of $C$ itself
\begin{equation}\label{eq:paramShift}
    \frac{\partial C(\theta)}{\partial \theta} = \frac{1}{2}\left( C(\theta + \tfrac{\pi}{4r}) - C(\theta - \tfrac{\pi}{4r}) \right).
\end{equation}
The requirement of the Hamiltonian to have only two eigenvalues is not restrictive for practical variational circuits: all single-qubit rotation gates and all Pauli operators have spectra $\{-1,1\}$, and tensor products of Pauli matrices likewise have only two eigenvalues. Although generalized shift rules exist for generators with larger spectra,
 the two-eigenvalue case above covers the vast majority of parametrized gates used in contemporary variational algorithms and forms the foundation of efficient gradient evaluation on near-term quantum hardware. See \cite{schuld2021machine} for a general treatment.
 
Variational quantum circuits have emerged as one of the most promising frameworks for achieving quantum advantage on NISQ devices. They are central to algorithms such as the Variational Quantum Eigensolver (VQE), the QAOA, and a wide range of methods in quantum machine learning. A comprehensive overview of variational techniques, their applications, and their main challenges can be found in the excellent survey by~\cite{cerezo2021variational}.

\subsection{Mixed states, quantum probability and Born's rule}
\label{sec:mixed_states}

In realistic scenarios, one rarely has complete knowledge of the pure state of the system. Noise, decoherence, or incomplete state preparation introduce additional layers of uncertainty.  To model this, quantum mechanics extends the notion of state to \emph{mixed states}, described not by a single vector but by a \textit{density operator} (or density matrix)
\begin{equation}\label{eq:mixedstate}
\rho = \sum_i p_i \ket{\psi_i}\bra{\psi_i}, 
\quad p_i \geq 0, \quad \sum_i p_i = 1,
\end{equation}
which represents a statistical ensemble of pure states $\{\ket{\psi_i}\}$.  Any pure state $\ket{\psi}$ can be thought as a rank-one mixed state $\rho = \ket{\psi}\bra{\psi}$. In fact, a mixed state $\rho$ is pure if and only if $\tr(\rho^2) = 1$. 

Although each coefficient $p_i$ in Eq.~\eqref{eq:mixedstate} is usually called the ``probability of $\rho$ being prepared in state $\ket{\psi_i}$'', the list $\{p_i\}$ should, in general, not be interpreted as a probability distribution. The reason is similar to that regarding the amplitudes of pure states: the representation Eq.~\eqref{eq:mixedstate} is not unique, it depends on the states $\{\ket{\psi_i}\}$ of the ensemble representation. 
For example, the simple state $\rho=\frac{1}{2}I \in \Hcal_1$ can be written in two very different ensembles:
\begin{equation}
    \rho = \frac{1}{2} \ket{0}\bra{0} + \frac{1}{2} \ket{1}\bra{1} = \frac{1}{2} \ket{+}\bra{+} + \frac{1}{2} \ket{-}\bra{-}.
\end{equation}
This, in practice, means that $\rho$ can be prepared from different mixtures. In conclusion, mixed states are not associated with one ensemble, only with the density operator itself. 

A more detailed way of representing mixed states is with respect to quantum bases. Let $\rho$ be as in Eq.~\eqref{eq:mixedstate},  and let $\mathcal{B} = \{\ket{\phi}_i\}_{i=1}^{2^n}$ be an arbitrary orthonormal basis for $\mathcal{H}_n$. Expanding each as $\ket{\psi_i} = \sum_j c_{i,j} \ket{\phi_j}$ and replacing in Eq.~\eqref{eq:mixedstate}, gives
\begin{equation}
\begin{split}
    \rho &= \sum_{i=1}^{2^n} p_i \left(\sum_{j=1}^{2^n} c_{i,j} \ket{\phi_j}\right)
    \left(\sum_{k=1}^{2^n} \bar{c}_{i,k} \bra{\phi_k}\right) \\
    &= \sum_{j,k=1}^{2^n} \left(\sum_{i=1}^{2^n} p_i c_{i,j} \bar{c}_{i,k} \right) \ket{\phi_j}\bra{\phi_k} 
    = \sum_{j,k=1}^{2^n} \rho_{j,k} \ket{\phi_j}\bra{\phi_k}.
\end{split}
\end{equation}
The diagonal elements $\rho_{j,j}$ are nonnegative and sum to one (because $\tr(\rho)$ is base-invariant) but do not constitute probabilities intrinsic to $\rho$ either. We will see that they are the probabilities of the outcomes of measuring $\rho$ in the basis $\mathcal{B}$. The off-diagonal elements, in turn, capture quantum coherences: they quantify the potential for interference effects in that basis. 
 
Finally, since a density operator $\rho$ is a positive semidefinite Hermitian operator on $\mathcal{H}_n$ that has trace equal to one, the spectral theorem implies that $\rho$ can be \textit{diagonalized} as 
\begin{equation}\label{eq:rho_diagonal}
    \rho = \sum_{i=1}^{2^n} \lambda_i \ket{v_i}\bra{v_i}, \quad \lambda_i = \braket{v_i | \rho | v_i}
\end{equation}
for some non-negative \textit{eigenvalues} $\lambda_i \geq 0$ and \textit{eigenvectors} $\ket{v_i}$ that satisfy $\rho \ket{v_i} = \lambda_i \ket{v_i}$, and $\sum_{i=1}^{2^n} \lambda_i = 1$ because $\tr(\rho) = 1$. If all the eigenvalues are distinct (non-degenerate) the spectral decomposition is unique up to phase factors in the eigenvectors, thus providing a unique probability distribution intrinsic to~$\rho$.

The algebraic operation for the action of a quantum operator $A:\Hcal_n \to \Hcal_n$ on a mixed state $\rho$, can be deduced as follows. Suppose $\rho$ is the ensemble in Eq.~\eqref{eq:mixedstate}, then the action of $A$ over $\rho$ is
\begin{equation}\label{eq:ArhoA}
    \sum_i p_i (A \ket{\psi_i})(A \ket{\psi_i})^{\dag} 
    = \sum_i p_i A \ket{\psi_i} \bra{\psi_i} A^{\dag} = A \rho A^\dag.
\end{equation}
Note however that $A \rho A^\dag$ is not necessarily a quantum state: it is positive, but $\tr(A \rho A^\dag) \leq 1$, with equality if and only if~$A$ is a unitary.  That is, the map $\rho \mapsto A \rho A^\dag$ is not \textit{trace preserving}. We will see in \sectionref{sec:Luders} how to compute updates to the state associated to measurements stemming from observables.

\subsubsection{Quantum probabilities}
\label{sec:quantum_probs}

Mixed states are the key concept in the theory of \textit{quantum probability} as it applies to quantum information science. In general terms, quantum probability can be understood as the non-commutative analog of Kolmogorov's axiomatic theory of probability. Quantum probability is rich in mathematical, physical and philosophical subtleties, ranging from the interpretation of the act of measurement and the meaning of events, to the structure of operator algebra and operator logic. Here, we will restrict ourselves to an informal understanding of quantum probability, focusing on the finite-dimensional case relevant to quantum computing, showcasing its contrast with classical probability theory and its connection with uncertainty quantification. For a more thorough treatment we refer the reader to the classical work of~\cite{accardi1982foundations}, a more modern treatment in~\cite{khrennikov2016probability},~\cite{meyer1995quantum} for probabilists and~\cite{wang2022quantum} for statisticians.

In finite-dimensional quantum probability, the \textit{probability space} is defined by the quantum state. Hence the use of the subindex $\rho$ of $\P_\rho$ in Eq.~\eqref{eq:probMx}: it means the probability ``in the state $\rho$'', or ``with respect to state $\rho$''. Given a mixed state $\rho \in \mathcal{H}_n$, the \textit{sample space} for $\P_{\rho}$ is the whole Hilbert space $\mathcal{H}_n$ itself. The role of \textit{random variables} is played by the observable operators. The \textit{events} correspond to the possible eigenspaces of such observables, and the value of the observable in such an event is the corresponding eigenvalue.

In classical probability theory \textit{indicator random variables} are used to decompose random variables and define the connection between expectation and probability. This role is played in quantum probability by the projector operators. Let $P$ be such an operator.  Since $P^2=P$, its eigenvalues are $0$ and $1$, which correspond the two outcomes of a binary experiment associated $P$. The possible events are the orthogonal subspaces $\text{Ran}(P)$ and $\text{Null}(P)$ of $\mathcal{H}_n$. For a state $\rho$, the probability $\P_\rho$ of $\text{Ran}(P)$ measures the fraction of $\rho$ in $\text{Ran}(P)$. If $\rho = \ket{\psi}\bra{\psi}$ is pure and $P = P_0$ as in Eq.~\eqref{eq:probMx}, then the ``fraction'' of $\rho$ along $\ket{0}$ is precisely 
\begin{equation}\label{eq:Prho1}
    \P_\rho(\text{observe } 0) = |\braket{0|\psi}|^2 = \braket{\psi|P_0|\psi} = \tr(P_0 \rho).
\end{equation}
If $\rho = \sum_{i} p_i \ket{\phi_i}\bra{\phi_i}$, then the quantum probability associated to $P_0$ on $\rho$ can be expanded in a sort of ``total probability formula'' as
\begin{equation}\label{eq:Prho2}
    \P_\rho(\text{observe }0) = \sum_i p_i \P_{\ket{\phi_i}\bra{\phi_i}}(\text{observe } 0) = \sum_i p_i \tr\left(P_0 \ket{\phi_i}\bra{\phi_i}\right) = \tr(P_0 \rho). 
\end{equation}
This formula involving the trace of the operator is called \textit{Born's rule}, and defines the way quantum probabilities and expectations are computed within each quantum state. We expand on it in the next section.

The fundamental difference between quantum and classical probability is that observables in quantum computing may not commute. Mathematically, this follows from the projector operators not forming a $\sigma$-algebra as in classical probability, but an orthomodular lattice where operations between events have different properties than the familiar set operations of classical probability. 

Consider for example the events corresponding to the subspaces spanned by the pure states $\ket{0}$ and $\ket{+} = \frac{1}{\sqrt{2}}(\ket{0} +\ket{1})$ in $\Hcal_1$.
Let $P_0=\ket{0}\bra{0}, P_+ = \ket{+}\bra{+}$ be their corresponding projectors.
To compute the probability of the ``intersection event'' of observing 0 \textit{and} +, or the ``joint distribution'' of $(P_0,P_+)$, one would look at the product of the projectors. However, it is easy to check that $P_0P_+ \neq P_+P_0$, so the product does not represent a well-defined joint observable. Physically, this means that there is no single probability for the simultaneous occurrence of both outcomes: measuring one observable disturbs the other, and the observed probabilities depend on the order of measurement.   

\subsubsection{Born's rule for quantum probabilities}\label{sec:Born}

In most textbooks of quantum computing, Born's rule  is a postulate connecting the abstract Hilbert space to experimental probabilities, but it can also be derived in more abstract settings. Here, we introduce it informally as an extension of the formulas \eqref{eq:Prho1} and \eqref{eq:Prho2} to general observables.

Suppose $O$ is an observable on $\Hcal_n$. Since it is Hermitian, its eigenvalues $\{o\}$ are real, and there is a collection $\{P_o\}$ of projectors onto the corresponding eigenspaces satisfying $\sum_o P_o = I$.
Note that, for convenience, we are using no index for the discrete set of eigenvalues, and then use the eigenvalue itself as an index for the corresponding projection, in contrast to the applied mathematics notation $o_j$  and $P_j$ for $j=1,\ldots,N$ with $N \leq 2^n$.
We say that $\{P_o\}$ is a \textit{Projective Measurement} or PVM for short; see \tableref{tab:operators}.
Given the system in state $\rho$, the \textit{quantum expectation} of $O$ with respect to $\rho$ can thus be computed as in classical probability by summing over all possible outcomes $o$ and multiplying by the corresponding quantum probability
\begin{equation}\label{eq:Etr}
    \EXP_{\rho}(O) = \sum_o o \, \P_{\rho}(\text{observe }o) = \sum_o o \tr(P_o \rho) = \tr(O\rho).
\end{equation}
In the special case of a pure state, this reduces $\EXP_{\ket{\psi}\bra{\psi}}(O) = \braket{\psi |O | \psi}$, and nicely connects to the cost functions of variational quantum circuits; see \sectionref{sec:VQC}.
A drawback of the PVM approach is that there can not be more than $2^n$ outcomes associated to each observable.

A more general setup corresponds to the \textit{Positive Operator-Valued Measure formalism}, or POVM for short. As we will in \sectionref{sec:Qchannels} these play an important role in the modelling of interactions between the system and the environment. A POVM is a collection of positive Hermitian operators $\{M_m\}$ such that $\sum_m M_m =I $. Each $M_m$ is called an \textit{effect operator}.
The labels $m$ are the possible outcomes of the measurement and the number of labels can be larger than $2^n$. These are not necessarily projectors, so a PVM is a special case of a POVM. At measurement, the probability of observing label $m$ of the POVM $\{M_m\}$ with respect to $\rho$ is computed as before,
\begin{equation}\label{eq:obsm}
    \P_{\rho}(\text{observe } m) = \tr(M_m \rho).
\end{equation}

Eq.~\eqref{eq:obsm} is called \textit{Born's rule} and is the more general notion of quantum probability used in quantum computing. It simplifies to Eq.~\eqref{eq:probMx} for the case of single-qubit pure states. In quantum computing, Born's rule encodes intrinsic quantum randomness, defines the statistical distribution of outcomes, determines the sampling complexity of algorithms, and sets the baseline against which all other uncertainties (noise, errors, model approximations) must be quantified. 

\subsubsection{L\"uders rule for post-measurement states and Kraus decomposition}
\label{sec:Luders}

As we mentioned in \sectionref{sec:meas_pure}, observing a  realization of a quantum system irreversibly changes the system itself. L\"uders rule specifies which state the system \textit{collapses} to after each observation.
For a PVM $\{P_o\}$ associated with an observable $O$ with eigenvalues $\{o\}$ as in Eq.~\eqref{eq:Etr}, L\"uders rule states that in a realization where the eigenvalue $o$ is observed, the system's state collapses to the event represented by the eigenspace $\text{Ran}(P_o)$. 

Suppose, for example, $\rho = \ket{\psi}\bra{\psi}$ which has spectrum $\{0,1\}$ so our PVM is simply $\{P_0,P_1\}$ in $\Hcal$ as in Eq.~\eqref{eq:Prho1}. If $0$ is observed, then the system collapses to $\text{Ran}(P_0)$, which is represented as a mixed state by $\ket{0}\bra{0}$. In terms of $P_0$ and the original state $\rho$, this post-measurement state can be obtained by taking the result of the projection $P_0 \rho P_0$ and normalizing by the probability of the event,
\begin{equation}
    \frac{P_0 \rho P_0 }{\tr(P_0 \rho)} = \frac{\ket{0}\braket{0|\psi}\braket{\psi|0}\bra{0}}{\braket{\psi|P_0|\psi}} = \frac{ \ket{0}\bra{0} \, |\bra{0|\psi}|^2}{|\bra{0|\psi}|^2} = \ket{0}\bra{0}.
\end{equation}
The above identity always holds for projections on one dimensional subspaces and $\rho$ need not be pure. For the general PVM $\{P_o\}$ on $\rho$, the post-measurement ``updated'' or ``collapsed'' state after observing $o$, is
\begin{equation}\label{eq:LudersPVM}
    \rho_o' = \frac{P_o \rho P_o}{\tr(P_o \rho)}.
\end{equation}
L\"uders rule in Eq.~\eqref{eq:LudersPVM} can be interpreted as an update yielding the intermediate mixed state
\begin{equation}
    \sum_o P_o \rho P_o
   = \sum_o \tr(P_o \rho) \rho_o'
\end{equation}
together with a Bayesian update using the observed value of $o$,
and the collapse onto the event associated to $P_o$.
The numerator contains the states on the range of $P_o$ under $\rho$, and the denominator normalizes by the probability of $P_o$.  The normalization in Eq.~\eqref{eq:LudersPVM} is necessary because projectors are not unitary. Thus, as pointed out after Eq.~\eqref{eq:ArhoA}, the \textit{unnormalized state} $P_o \rho P_o$ is not a true quantum state.

L\"uders' rule in the projective form of Eq.~\eqref{eq:LudersPVM} cannot be directly extended to a general POVM $\{M_m\}$ by simply replacing the projectors with POVM elements. The probability of the observing the outcome $m$ is given by $\P_{\rho}(\text{observe } m) = \tr(M_m \rho)$ as in Eq.~\eqref{eq:Prho2}. But, while for the PVM in Eq.~\eqref{eq:LudersPVM}, $\tr(P_o \rho) = \tr(P_o^2 \rho)= \tr(P_o \rho Po)$ correctly normalizes $\rho_o'$, we have $\tr(M_m \rho) \neq \tr(M_m \rho M_m)$ so the expression $M_m\rho M_m$ does not in general lead to the correct normalization. 

To describe both the measurement statistics and the corresponding state update, one introduces a family of completely positive maps $\{\mathcal I_m\}$, called a \textit{quantum instrument}, such that
\begin{equation}\label{eq:LudersPOVM}
    \rho \mapsto \rho'_m = \frac{\mathcal{I}_m(\rho)}{\tr(M_m \rho)} \quad \text{and} \quad \tr(\mathcal{I}_m(\rho))=\mathbb{P}_\rho(\text{observe }m)=\tr(M_m\rho).
\end{equation}
Thus $\mathcal I_m(\rho)$ is a subnormalized post-measurement state associated with outcome $m$, which by analogy with the classical case, represents the occurrence of event $m$ under $\rho$.

A canonical choice of instrument is obtained from the positive square root of $M_m$. Since each POVM element $M_m$ is positive semidefinite, it admits a unique positive square root $\sqrt{M_m}$ satisfying $M_m = \sqrt{M_m}^{\,\dagger}\sqrt{M_m}.$ This yields the L\"uders instrument
$\mathcal I_m(\rho)=\sqrt{M_m}\,\rho\,\sqrt{M_m}$ for which
$\tr\!\bigl(\sqrt{M_m}\,\rho\,\sqrt{M_m}\bigr)=\tr(M_m\rho)$
as required.

The decomposition $M_m = \sqrt{M_m}^\dag\sqrt{M_m}$ is, however, not the only method for obtaining an instrument. In general, we can decompose $M_m$ into its \textit{Kraus operators} $\{K_{m,i}\}_i$, such that
\begin{equation}\label{eq:KrauseDec}
    M_m = \sum_i K^\dag_{m,i} K_{m,i}, \quad \sum_m M_m = \sum_{i,m} K^\dag_{m,i} K_{m,i} = I.
\end{equation}
Eq.~\eqref{eq:KrauseDec} is called the \textit{Kraus decompostion} or \textit{operator-sum representation} of the POVM element. Each term $K_{m,i}^\dag K_{m,i}$ is called a ``partial effect'' and is interpreted as a microscopic transformation leading to the macroscopic outcome $M_m$. Any choice of Kraus decomposition will yield the same statistics on the observable, but produces a different post-measurement state, namely the state defined by the instrument
\begin{equation}\label{eq:postmeasPOVM}
    \mathcal{I}_m(\rho) = \sum_i K_{m,i} \, \rho \, K_{m,i}^{\dag}, \quad \text{and} \quad
    \rho_m' = \frac{\mathcal{I}_m(\rho)}{\tr(M_m \rho)}.
\end{equation}
The first identity in Eq.~\eqref{eq:postmeasPOVM} is the Kraus decomposition of the instrument $\mathcal{I}_m$; the second expresses the post-measurement state for a general POVM in form reminiscent of a \textit{law of total probability} over all partial effects.

In practice, the Kraus decomposition is determined by the physical architecture of the measurement device. The corresponding instruments are implemented in quantum circuits by introducing ancilla systems and applying suitable joint unitaries before measuring the ancilla registers. The mathematical foundation of this construction is provided by the Stinespring dilation theorem, which states that any completely positive map can be realized as a unitary interaction on a larger Hilbert space followed by a partial measurement or trace over an ancillary system; see \cite{watrous2018theory} for details.

\subsection{Mixed states in quantum circuits}

While pure states provide an idealized description of quantum information flow, practical quantum circuits must account for mixed states. In circuit language, the density operator $\rho$ evolves through quantum gates via the action of a unitary $\rho \mapsto U \rho U^\dag$, and measurement operators as described in \sectionref{sec:Born} and \sectionref{sec:Luders}. Mixed-state representations allow us to describe such processes compactly as completely positive maps, track subsystem dynamics using the partial trace, and model imperfect operations through coupling with an extended Hilbert space. In this section we review composite-system descriptions and their reduced states, show how purification provides a constructive link between noisy and ideal evolutions, and outline how quasiprobability representations express circuit-level uncertainty in a form directly comparable to classical probabilistic models.

\subsubsection{Composite systems, partial trace and purification}\label{sec:compSystems}

Density operators are used to describe composite quantum systems, composed for example of a state in a quantum circuit and the surrounding environment. For these we may use a product Hilbert space $\mathcal{H}_{AB} = \mathcal{H}_A \otimes \mathcal{H}_B$ where $A,B$ label the different subsystems. A mixed state in this system $\rho_{AB}$ is, for most practical cases entangled and therefore not a tensor product of states in each system. However, the composite state can be reduced back to either system using the partial trace. The \textit{reduced states} in each system are obtained by ``tracing out'' the other:
\begin{equation}\label{eq:redstate}
    \rho_A = \tr_B(\rho_{AB}), \quad \rho_B = \tr_A(\rho_{AB}).
\end{equation}
The partial trace is one of the most useful operations in quantum computing and plays the role of marginalization in probability theory. It is defined on products of pure states as follows: consider arbitrary states $\ket{a_1},\ket{a_2} \in \mathcal{H}_A$ and $\ket{b_1},\ket{b_2} \in \mathcal{H}_B$, and define $\rho_{AB} = \ket{a_1}\bra{a_2} \otimes \ket{b_1}\bra{b_2}$, then
\begin{equation}\label{eq:traceprop}
    \rho_A =\tr_B(\rho_{AB}) = \ket{a_1}\bra{a_2} \tr(\ket{b_1}\bra{b_2})
    = \ket{a_1}\bra{a_2} \,\braket{b_1|b_2}.
\end{equation}
For an arbitrary pure state $\ket{\psi} \in \Hcal_{AB}$, the partial trace is extended by linearity and can be deduced using Schmidt's decomposition; see Eq.~\eqref{eq:SchmidtDec}. Specifically, decompose $\ket{\psi} = \sum_i \lambda_i \ket{\psi_{A,i}} \otimes \ket{\psi_{B,i}}$. Then tracing out $B$ produces the reduced state 
\[
\rho_A = \tr_B(\ket{\psi}\bra{\psi}) = \sum_i \lambda^2_i \ket{\psi_{A,i}}\bra{\psi_{A,i}} \in \Hcal_A\otimes \Hcal_A^*.
\]

In addition to the partial trace, in a similar way, we also define a way to partially apply vectors from a subsystem Hilbert space, say \(\Hcal_B\), to operators defined on the product space. First, we consider a rank-one operator \(U = 
(\ket{a_1} \otimes \ket{b_1}) (\bra{a_2} \otimes \bra{b_2}) =
\ket{a_1}\bra{a_2} \otimes \ket{b_1}\bra{b_2}\) as before. Then, for any \(\ket{\psi_{B,1}},\ket{\psi_{B,2}}\) the partial applications are defined as
\begin{equation*}
\begin{aligned}
\bra{\psi_{B,1}} U &= \braket{\psi_{B,1}|b_1}\,\ket{a_1}(\bra{a_2} \otimes \bra{b_2}), \\
\bra{\psi_{B,1}} U \ket{\psi_{B,2}} &= \braket{\psi_{B,1}|b_1}\braket{\psi_{B,2}|b_2}\,\ket{a_1}\bra{a_2}.
\end{aligned}
\end{equation*}
Again, since the rank-one operators form a basis of $(\Hcal_{A}\otimes\Hcal^*_{A})\otimes(\Hcal_{B}\otimes\Hcal^*_{B})\cong\Hcal_{AB}\otimes\Hcal_{AB}^*$ this can be extended by linearity to any \(U\in\mathcal{L}(\Hcal_{AB})\).
Furthermore, for a basis \(\{\ket{\psi_{B,i}}\}\) of \(\Hcal_{B}\) we have the useful formula
\begin{equation}
\label{eq:partial_app_trace}
\tr_B(U) = \sum_{i} \bra{\psi_{B,i}} U \ket{\psi_{B,i}},
\end{equation}
linking the partial applications and the partial trace.

To compute the expectation of an observable over a subsystem, we can use the partial trace to write the expectation in terms of only reduced systems. Suppose for example that we have a composite mixed state $\rho_{AB} \in \mathcal{H}_{AB}\otimes \mathcal{H}_{AB}^*$ and we want to observe $O_A$ on the subsystem $A$. This is equivalent to observing $O_A \otimes I$ on $\rho_{AB}$. The quantum expectation can be written with respect to the composite system as in Eq.~\eqref{eq:Etr}, or using the partial trace, first tracing out $B$, and then taking trace over $A$, 
\begin{equation}
    \EXP_{\rho_{AB}}(O_A) = \EXP_{\rho_{AB}}(O_A \otimes I) = \tr_{AB}(\rho_{AB}(O_A \otimes I)) = \tr_A(\tr_B(\rho_{AB}) O_A) = \tr_A(O_A \rho_A).
\end{equation}

Just like the distribution of random variable can be always obtained by marginalization of a joint distribution, any mixed state can be obtained by tracing out a larger pure state. This is called \textit{purification} and works as follows. Suppose $\rho$ is a mixed state of the system $\mathcal{H}_n$.
Then it is possible to introduce a ``reference'' system $\mathcal{R}$ such that there is a pure state $\ket{\psi_\rho} \in \mathcal{H}_n \otimes \mathcal{R}$ and $\rho = \tr_{\mathcal{R}}(\ket{\psi_\rho}\bra{\psi_\rho})$. This pure state is not unique but can always be constructed using $\mathcal{R} = \mathcal{H}_n$: start by diagonalizing $\rho = \sum_{i} \lambda_{i} \ket{v_i}\bra{v_i}$ and define the pure state on $\mathcal{H}_n \otimes \mathcal{H}_n$ with Schmidt decomposition $\ket{\psi_\rho} = \sum_i \sqrt{\lambda_i} \ket{v_i}\otimes \ket{v_i}$.

The purification of a quantum state provides an important interpretation of uncertainty arising from interaction with an environment. Let the reference system $\mathcal{R}$ represent this environment. If $\rho$ is associated to the pure state $\ket{\phi}$, then $\ket{\psi_\rho}$ can be chosen as $\ket{\psi_\rho} = \ket{\phi} \otimes \ket{e}$ for any $\ket{e} \in \mathcal{R}$; in this case there is no entanglement between the system and the environment. Conversely, if $\rho$ is mixed, then every purification $\ket{\psi_\rho}$ has Schmidt rank greater than one, and is thus entangled across the system-environment partition. In general, the degree of ``mixedness'' of $\rho$, and thus the amount of uncertainty in this mixed state induced by the environment, can be quantified by the amount of entanglement present in the corresponding purified state.

\subsubsection{Quasiprobability representation}

Quasiprobability representations in quantum mechanics date back to the work of~\cite{wigner1932quantum} that assigned to every quantum state a real-valued distribution reproducing quantum expectation values in analogy with classical statistical mechanics, but at the cost of allowing negative values. A modern formulation of quasiprobability representations can be found in~\cite{ferrie2009framed,ferrie2011quasi}. It provides a way of representing any mixed state and observable as a random variable whose distribution is given by quasiprobabilities: real numbers not neccessarily in $[0,1]$ but with properties resembling those of classical probability. 

The mathematical formulation uses the concept of a \textit{frame} of operators for the ambient Hilbert space $\mathcal{H}_n$. A frame is a finite spanning family $\{F_\lambda\}$ of operators into which any Hermitian operator can be written as a linear combination. The coefficients in the expansion are computed from a \textit{dual} frame $\{G_\lambda\}$. Specifically, any mixed state $\rho$ can be written as 
\begin{equation}\label{eq:repFrames}
    \rho = \sum_{\lambda} \tr(G_\lambda \rho) F_\lambda.
\end{equation}
The fact that we use frames and not common bases means that the number of $\lambda$s may be larger than $2^{2n}$ -- the dimension of the space of operators -- and the representation Eq.~\eqref{eq:repFrames} may not be unique. It is common to take a ``tight'' frame in which $\sum_{\lambda} G_\lambda = I$ and $\sum_\lambda F_\lambda \propto I$. In this setting the real numbers $\{\tr(G_\lambda \rho)\}$ are not in $[0,1]$ but satisfy $\sum_\lambda \tr(G_\lambda \rho) = 1$, and hence they are called the \textit{quasiprobability distribution} of $\rho$ with respect to the frame $G$. 

Suppose $O$ is an observable (a Hermitian operator) and we want to compute its expectation on the output $\rho_\tout = U \rho U^\dag$ of a quantum circuit. The operator $O$ can be written as in Eq.~\eqref{eq:repFrames}, or exchanging the roles of $F$ and $G$ as $O = \sum_{\lambda'} \tr(O F_{\lambda'}) G_{\lambda'}$. The quantum expectation of the observable on $\rho_{\tout}$ is then 
\begin{equation}\label{eq:totalQP}
    \EXP_{\rho_\tout}(O) = \sum_{\lambda,\lambda'} \tr(O F_{\lambda'}) \tr(G_{\lambda'} U F_{\lambda} U^\dag) \tr(G_\lambda \rho)
\end{equation}
which is a type of total quasi-probability formula in terms of the ``conditional'' distribution $\tr(G_{\lambda'} U F_{\lambda} U^\dag)$ of $\lambda'$ given $\lambda$, encoding the action of the unitary $U$~\cite{pashayan2015estimating}. 

We will see in \sectionref{sec:UQmethods} that quasiprobability representations have become a powerful tool in quantum computing, enabling efficient simulation of restricted circuit classes, rigorous quantification of computational resources, and practical error mitigation strategies on noisy intermediate-scale quantum (NISQ) devices.

\subsection{Quantum channels, noise and errors}
\label{sec:Qchannels}

The model used for the most general transformation that a quantum state can undergo, encompassing both unitary dynamics and measurement, as well as noise and open-system effects, is the \textit{quantum operation}, also called \textit{quantum channel}. In general, a quantum channel is a linear map 
\begin{equation}
    \rho \mapsto \mathcal{E}(\rho)\label{eq:quantum_channel}
\end{equation}
that maps mixed states to mixed states (possibly on a different quantum system). Mathematically, the defining conditions for a quantum channel are: (i) $0 \leq \tr(\mathcal{E}(\rho)) \leq 1$ for all $\rho$, and (ii) $\mathcal{E}$ is completely positive, namely for any $k$, the extended map $\mathcal{E} \otimes I^{\otimes k}$ sends positive semidefinite operators to positive semidefinite operators.

The first condition on $\mathcal{E}$ is the analog of requiring stochasticity or substochasticity on a Markovian transition operator. Those quantum operations with $\tr(\mathcal{E}(\rho)) = 1$ for all $\rho$ are called completely positive trace preserving (CPTP) maps, and correspond to the transformation of quantum states in which no measurement takes place, for example by the action of a unitary $\mathcal{E}(\rho) = U \rho U^\dag$. The case of $\tr(\mathcal{E}(\rho)) < 1$ corresponds to the transformation that a state undergoes after a measurement, for example given by each component of an instrument as in Eq.~\eqref{eq:LudersPOVM}. 

The complete positivity condition is particularly important in the context of noise, and generally of composite quantum systems. It essentially says that, if $\mathcal{E}$ is a channel that acts only on a subsystem $A$ and we extend it to $\mathcal{E} \otimes I_B$ on a composite system $\mathcal{H}_{AB}$, then $(\mathcal{E} \otimes I_B)(\rho_{AB})$ will still be a valid mixed state for any composite state $\rho_{AB}$. This requirement ensures that the map $\mathcal{E}$ can be consistently applied to parts of entangled systems without producing nonphysical states with negative eigenvalues. As we saw in \sectionref{sec:Luders} for the case of instruments, completely positivity implies a \textit{operator-sum} or \textit{Kraus decomposition} of the channel,
\begin{equation}
    \mathcal{E}(\rho) = \sum_i K_i \rho K_i^\dagger, 
    \qquad 
    \sum_i K_i^\dagger K_i = I,
\end{equation}
which provides a compact and universal representation of all physical quantum evolutions. The Kraus operators $K_i$ describe the different microscopic transformations that the system may undergo, and the completeness relation guarantees that the overall process preserves total probability.

Quantum operations can be used to describe all physically realizable evolutions of quantum systems, including both unitary (noise-free) and non-unitary (open, noisy) dynamics. They are the language in which we can describe \textit{quantum noise} and \textit{quantum error}. Specifically the term quantum noise refers to a physical process that produces uncontrolled, typically small, perturbations affecting quantum systems during storage, evolution, or measurement. Quantum errors are the undesired effects of the noise and we will talk about them in \sectionref{sec:error_corr}.

\begin{figure}[htbp]
    \centering
    \includegraphics[width=0.45\linewidth]{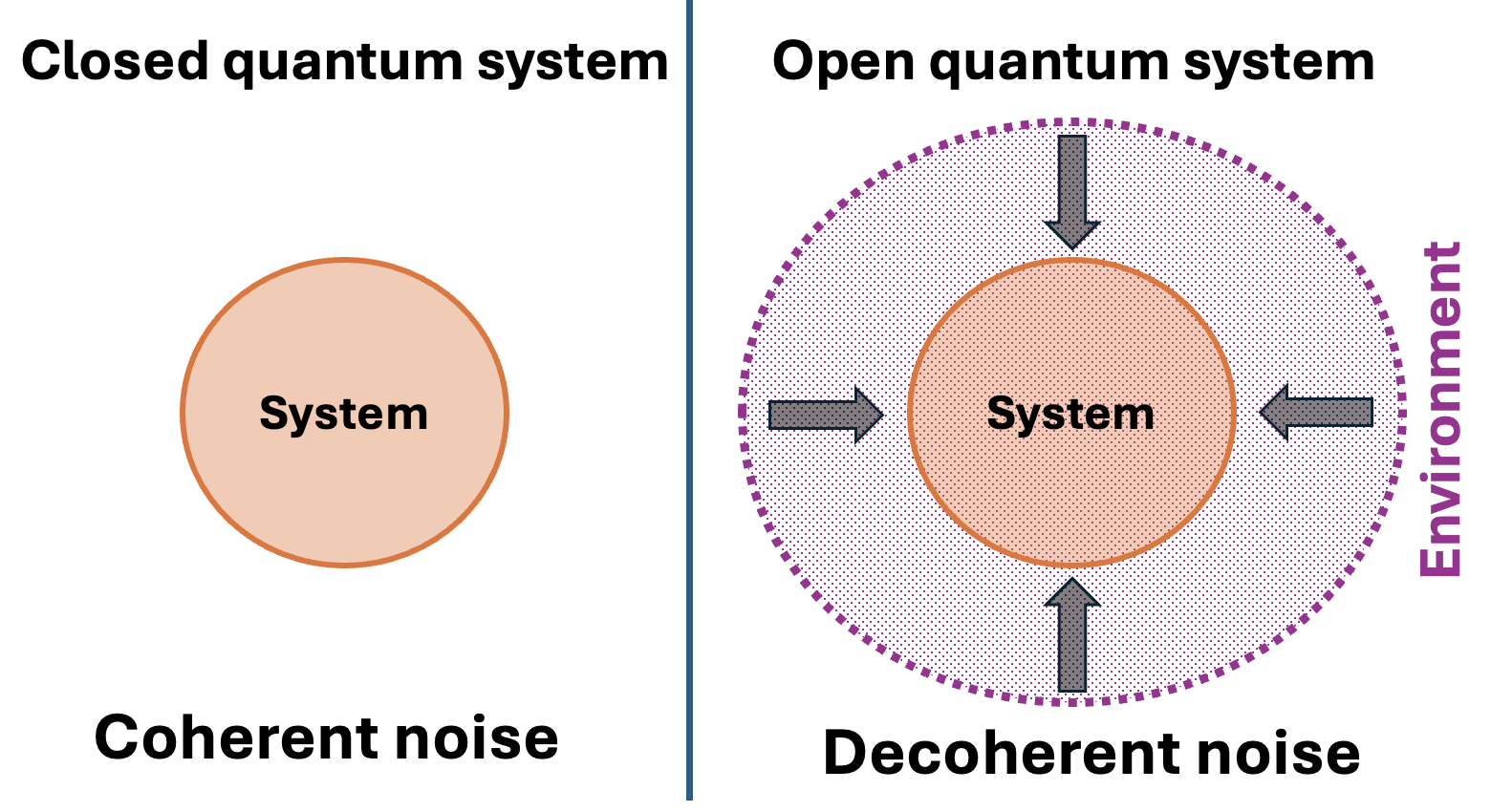}
    \hspace{1em}
    \includegraphics[width=0.5\linewidth]{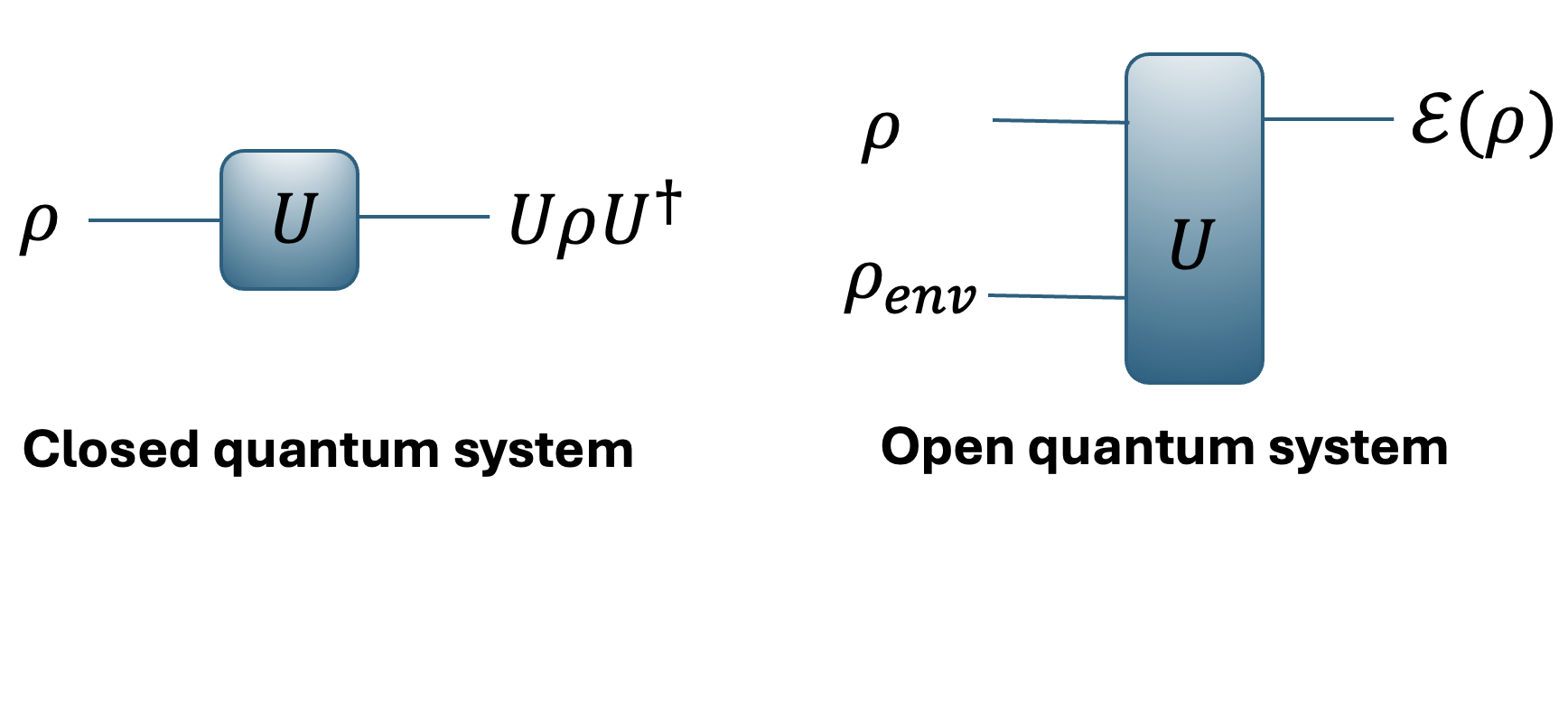}
    \caption{Schematic representation of the open and closed open quantum systems. }
    \label{fig:enter-label}
\end{figure}

Quantum noise is modeled by a quantum channel $\mathcal{E}$ and can be of two different types: \textit{coherent noise} and \textit{decoherent noise}. The first type arises in a \textbf{closed quantum system} from systematic, unitary miscalibrations of the intended quantum gates and it creates an undesired  drift, a type of \textit{epistemic uncertainty}. In contrast, \textit{decoherent noise} occurs in an {\bf{open quantum system}}~\cite{lidar2019lecture}, where the interaction with the environment induces stochastic processes such as relaxation or dephasing, analogous to noise in a classical communication channel, and gives rise to \textit{aleatoric uncertainty}.
Mathematically, both types of noise are described within the general framework of CPTP maps: coherent noise corresponds to purely unitary channels, while decoherent noise requires more general representations such as Kraus operators and Lindblad master equations, as described in the next section.

Specifically, if we denote by $U$ the intended unitary operation that the circuit is designed to implement, the corresponding coherent noise channel can be modeled as the CPTP map
\begin{equation}\label{eq:Ecoh}
    \mathcal{E}_H(\rho) = e^{-i \epsilon H} U \rho U^{\dag} e^{i \epsilon H},
\end{equation}
where $H$ is a Hamiltonian and $\epsilon>0$ is small. This is equivalent to the channel corresponding to the perturbed unitary $\tilde{U} = e^{-i \epsilon H} U$. Examples of coherent noise in quantum circuits include: over/under rotations, phase misalignments and cross-talk between qubits. Their net effect is a drift away from the intended quantum operation that might accumulate with idle time or circuit depth. These cannot be ``averaged out'' and must be corrected. Most \textit{quantum error-correction} techniques deal with this type of noise.

\subsection{The environment and decoherent noise}\label{sec:env_noise}

Decoherent noise arises because real quantum algorithms are open quantum system interacting with an uncontrollable, unknown environment. This type of noise cannot be modeled by the action of a unitary on solely the circuit, and is therefore irreversible. 

Mathematically, the interaction between circuit and environment can be modeled in a Hilbert space $\mathcal{H} = \mathcal{H}_n \otimes \mathcal{H}_E $ composed of a \textit{principal system} $\mathcal{H}_n$ where the circuit gates operate, and an \textit{environment} $\mathcal{H}_E$. In the simplest case, a computational state
$\rho_C \in \mathcal{H}_n \otimes \mathcal{H}_n^*$
is carefully prepared as to eliminate all its correlations with the environment, and the circuit calculations starts with a separable state $\rho_C \otimes \rho_E$. The intended circuit computation is a unitary $U_C$ acting on $\mathcal{H}_n$, but the evolution of the whole state is a unitary $U$ on $\mathcal{H}$. The ``real'' transformation in $\Hcal_n$ is the quantum operation defined by tracing out the environment
\begin{equation}\label{eq:ErhoC}
    \mathcal{E}(\rho_C) = \tr_E(U(\rho_C \otimes \rho_E)U^\dag)
\end{equation}
which is called the \textit{dilation formula}.

Suppose the environment starts in the pure state 
\(\rho_E = \ket{e}\bra{e}\).
and let $\{\ket{e_k}\}$ be an orthonormal basis for the environment Hilbert space $\mathcal{H}_E$. 
Combining the partial trace in Eq.~\eqref{eq:ErhoC} with the representation through partial application from~\eqref{eq:partial_app_trace} we obtain
\begin{equation}\label{eq:kraus}
\mathcal{E}(\rho_C)
= \sum_k \bra{e_k} \, U \bigl(\rho_C \otimes \ket{e}\bra{e} \bigr) U^\dag \, \ket{e_k}
= \sum_k E_k \, \rho_C \, E_k^\dagger,
\end{equation}
where we have defined Kraus operators as the partial application
\[
E_k := \bra{e_k} U \ket{e} \in \mathcal{H}_n \otimes \mathcal{H}_n^*.
\]
This is called the \emph{Kraus representation} of the channel.

If we now measure (or let the system interact with) the environment on the basis $\{\ket{e_k}\}$, then the principal state will collapse to a state proportional to $\tr_E(\ket{e_k}\bra{e_k} U(\rho_C \otimes \rho_E)U^\dag \ket{e_k}\bra{e_k}) = E_k \rho_C E_k^\dag$. Namely to the state
\begin{equation}
    \rho'_{C_k} = \frac{E_k \rho_C E_k^\dag}{\tr(E_k \rho_C E_k^\dag)}.
\end{equation}
As in Eq.~\eqref{eq:LudersPOVM}, the probability of observing $k$ is the same as the normalization factor, and hence we can write the full channel as
\begin{equation}\label{eq:ErhoCprob}
    \mathcal{E}(\rho_C) = \sum_k \rho'_{C_k} \, \P_{\rho_C}(\text{observe } k).
\end{equation}
The quantum channel can thus be thought as a sort expectation over the possible principal mixed states that result from the environmental interaction. 

If the Kraus decomposition of $\mathcal{E}$ has more than one $E_k$, then we see from Eq.~\eqref{eq:ErhoCprob} that the channel $\mathcal{E}$ is truly stochastic and similar to a noisy communications channel: the information we retrieve from this quantum operation and measurement is the state $\rho'_{C_k}$ with probability $\tr(E_k \rho_C E_k^\dag)$. This is what defines decoherent noise channels. We can also formally understand them by the dilation formula in Eq.~\eqref{eq:ErhoC}: the partial tracing out of the $E$ system makes the channel $\mathcal{E}$ non-reversible, introducing the environmental degrees of freedom in the $E_k$. This \textit{extrinsic uncertainty} is precisely modeled by the probabilistic mixture of operations in Eq.~\eqref{eq:ErhoC}. It is not born out of lack of knowledge of $\rho_C$, but comes from tracing out the unknowable $E$.

In the case of coherent noise shown in Eq.~\eqref{eq:Ecoh}, $E_H$ is already in Kraus form with a single $K = e^{-i \epsilon H} U$ and is thus deterministic. Even for the composite system  $\mathcal{H} = \mathcal{H}_n \otimes \mathcal{H}_E $, if the decoherence affects only on the principal subsystem, then we can write it composite unitary as $U = e^{-i \epsilon H}U_C \otimes I_E$. In this case the principal state does not get entangled with the environment and, using Eq.~\eqref{eq:traceprop} we see that tracing out the environment has no effect,
\begin{equation*}
    \tr_E\left(U(\rho_C \otimes \rho_E)U^\dag \right) = 
    \tr_E\left( \left( e^{-i \epsilon H}\, U_C \, \rho_C \, U_C^\dag \, e^{i \epsilon  H}  \right) \otimes \rho_E \right) 
    =  e^{-i \epsilon H}\, U_C \, \rho_C \, U_C^\dag \, e^{i \epsilon  H}. 
\end{equation*}

In general, quantum noise is thus modeled specifying the environment state $\rho_E$ and the joint unitary $U$ in Eq.~\eqref{eq:ErhoC}. The canonical channels for modeling decoherent quantum noise are:
\begin{itemize}
    \item Bit-flip: Models random Pauli-$X$ errors with probability $p$.
    \item Phase-flip: Models random Pauli-$Z$ errors with probability $p$.
    \item Amplitude damping: Models energy dissipation (e.g.\ spontaneous emission) with probability $\gamma$.
    \item Phase damping: Models dephasing (loss of coherence without energy loss) with probability $p$.
    \item Depolarizing noise: With probability $p$, the state is replaced by the maximally mixed state.
\end{itemize}
Each can be specified by a parametric version of the Kraus expansion in Eq.~\eqref{eq:kraus}. Consider for example the noise channel for the bit-flip of the first qubit of a $n$-qubit circuit. Its probabilistic representation analog to Eq.~\eqref{eq:ErhoCprob} is given by the identity with probability $1-p$ and conjugation with the $X$ gate on the first qubit with probability $p$, that is,
\begin{equation}\label{eq:Ebitflip}
    \mathcal{E}_{\text{BF}1}(\rho_C; p) = (1-p) \rho_C + p \; X_1 \rho_C X_1, \quad X_1 = \left(\begin{array}{cc}
        0 & 1 \\
        1 & 0
    \end{array}\right) \otimes I_{n-1}
\end{equation}
The corresponding Kraus operators are $E_0 = \sqrt{1-p} I_n$, $E_1 = \sqrt{p} X_1$ so that $E_0E_0^\dag + E_1E_1^\dag = I_n$. A full representation in terms of the dilation formula Eq.~\eqref{eq:ErhoC} can be obtained by considering the environmental state $\rho_E = \ket{0}\bra{0}$ and the unitary $U =CX_1(I_n \otimes R_p)$ where $R_p$ is a rotation with probability $p$ on the environment, $R_p \ket{0} = \sqrt{1-p} \ket{0} + \sqrt{p}\ket{1}$, and $CX_1 = I_n \otimes \ket{0}\bra{0} + X_1 \otimes \ket{1}\bra{1}$ is a controlled gate that applies $X_1$ to the system only if the environment is in $\ket{1}$.

\subsection{Quantifying quantum errors and the fidelity}

Quantifying uncertainty in the context of quantum channels and noise involves measuring some sort of distance between the actual channel and the expected unitary behavior of a quantum circuit. For pure states, the fidelity is simply their simlarity as vectors in $\Hcal_n$, $F(\ket{\psi},\ket{\phi}) = |\braket{\psi|\phi}|^2$. For two mixed states $\rho$, $\sigma$, this definition extends to what is called \textit{Uhlman fidelity}
\begin{equation*}
    F(\rho,\sigma) = \left(\tr\sqrt{\sqrt{\rho} \, \sigma \, \sqrt{\rho}} \right)^2
\end{equation*}
Note that if $\rho,\sigma$ commute and we diagonalize them with respect to the same basis Eq.~\eqref{eq:mixedstate}, $\rho = \sum_i \lambda_i \ket{v_i}\bra{v_i}$, $\sigma = \sum_i \eta_i \ket{v_i}\bra{v_i}$, then $F(\rho,\sigma) = \tr\left(\sqrt{\rho}\sqrt{\sigma} \right)^2= (\sum_i \sqrt{\lambda_i \eta_i})^2$, which is the classical fidelity of the probability mass functions $\{p_i\}$, $\{q_i\}$. 

To quantify the level of noise produced by the environment in the quantum channel in Eq.~\eqref{eq:ErhoC} we can consider the fidelity between the intended channel $\mathcal{U}(\rho) = U \rho U^\dag$ and the ``real'' channel $\mathcal{E}(\rho)$, namely $F(\mathcal{U}(\rho),\mathcal{E}(\rho))$. Common approaches to consider the fidelity over a range of inputs include the \textit{worst-case gate fidelity}, which minimizes over all pure states (and hence over all states because $F(\rho,\sigma)$ is jointly concave)
\begin{equation*}
    F_{\min}(\mathcal{U},\mathcal{E})
    = \min_{\rho = \ket{\psi}\bra{\psi}} F(\mathcal{U}(\rho), \mathcal{E}(\rho))
    = \min_{\ket{\psi}} \braket{\psi| U^\dag \mathcal{E}(\ket{\psi}\bra{\psi}) U | \psi} ,
\end{equation*}
or the \textit{average gate fidelity} with respect to the Haar measure~\cite{mele2024introduction} ,the uniform probability measure over the unitary states,
\begin{equation*}
    F_{\text{avg}}(\mathcal{U},\mathcal{E}) = \int_{\mathcal{H}_n} \braket{\psi| U^\dag \mathcal{E}(\ket{\psi}\bra{\psi}) U | \psi} \text{dHaar}(\ket{\psi}).
\end{equation*}
In both cases of the fidelity, $0 \leq F(\mathcal{U},\mathcal{E}) \leq 1$ with $F(\mathcal{U},\mathcal{E})$ if and only if $\mathcal{E} = \mathcal{U}$. The \textit{infidelity} is defined as $r = 1-F$. For example, for the one-qubit bit-flip channel, the gate fidelities are
\begin{equation}\label{eq:F_bitflip}
    F_{\min}(\mathcal{U},\mathcal{E}_{\text{BF}1}) = 1-p,\quad 
    F_{\text{avg}}(\mathcal{U},\mathcal{E}_{\text{BF}1}) = 1 - \frac{2^n}{2^n +1}p.
\end{equation}

Using the fidelity we can see an important distinction between the accumulation of uncertainty in coherent and decoherent noise. Consider the application of $N$ bit-flip channels with fidelity given by Eq.~\eqref{eq:F_bitflip}. Then, as in the case of a simple random walk, the fidelity will scale as
\begin{equation*}
    F_{\min}(\mathcal{U},\mathcal{E}_{\text{BF}1}^N) \sim (1-p)^N = 1-Np + O(p^2).
\end{equation*}
The infidelity accumulates linearly with $Np$. For coherent noise, consider the fidelity between a desired pure state $\ket{\psi}$ and the effect of a small rotation applied $N$ times 
\begin{align*}
    \left |\Braket{\psi |e^{-i N \epsilon H} \psi} \right|^2 
    &= \left(1 - i N \epsilon \braket{\psi|H|\psi} - (N \epsilon)^2 \braket{\psi|H^2|\psi} + O((N\epsilon)^3)\right)^2\\
    &= 1- (N \epsilon)^2 \text{Var}_{\psi}(H) + O((N\epsilon)^4),
\end{align*}
where $\text{Var}_{\psi}(H) = \braket{\psi|H|\psi}^2 - \braket{\psi|H^2|\psi}$ is the quantum variance of $H$ on $\ket{\psi}$. The infidelity accumulates quadratically on $N \epsilon$. This difference underscores the problem with the presence of even small rates of coherent errors.

Another notion of fidelity stems from considering entanglement as an information-like resource in quantum computing. Here, we want to measure how well is a quantum channel preserving entanglement between the system and a reference (or environment) system. This leads to the concept of \textit{entanglement fidelity}. 

Consider a partition $\mathcal{H} = \mathcal{H}_n \otimes \mathcal{H}_E $ as in \sectionref{sec:env_noise} and a ``real'' quantum channel $\mathcal{E}(\rho_C)$ on the principal system as in Eq.~\eqref{eq:ErhoC}. The entanglement fidelity $F_\text{ent}(\rho,\mathcal{E})$ measures how much of the entanglement between the systems is preserved under the action of the channel. Or equivalently, how much of the mixedness of $\rho_C$ is mantained (see \sectionref{sec:compSystems}). It is defined by the fidelity in $\mathcal{H}$ between a purification of $\rho_C$ and its evolved state under the extended channel $I_R \otimes \mathcal{E}$,
\begin{equation*}
    F_\text{ent}(\rho,\mathcal{E}) = \braket{ \psi_{\rho_C} |\,(I_{\mathcal{R}} \otimes \mathcal{E}) (\ket{\psi_{\rho_C}}\bra{\psi_{\rho_C}}) \, | \psi_{\rho_C}}.
\end{equation*}
It can be shown that $0 \leq F_\text{ent}(\rho,\mathcal{E}) \leq1$ and its value does not depend on the purification of $\rho_C$. Values of the entanglement fidelity very close to one, indicate that most of the entanglement has been preserved.

\subsection{Quantum noise as a time-continuous process}
\label{sec:LME}

We have so far, represented noise as a sequence of discrete quantum channels acting on a state after each gate or circuit layer. This framework is natural for circuit models but does not reflect the fact that in physical devices, interactions with the environment occur continuously in time. To capture this more realistic setting, one introduces the \textit{Lindblad formalism} or \textit{master equation theory}~\cite{manzano2020LME}, which characterizes the infinitesimal time evolution of an open quantum system under the assumptions of weak coupling and Markovianity.

From the continuous-time perspective, an isolated quantum system with Hamiltonian $H$ is modeled by the evolution of a mixed state $\rho(t)$ via the von Neumann equation
\begin{equation}
    \deriv{\rho(t)}{t} = -i \left[H,\rho(t)\right],
\end{equation}
where $[A,B] = AB - BA$ denotes the commutator between operators. 
This describes unitary evolution: the system remains perfectly coherent and no information is lost to the outside world. However, realistic devices are never perfectly isolated; their interaction with an environment turns them into open quantum systems and introduces irreversible processes that must be accounted for. In this case, an open system model introduces the \textit{Lindblad operators} $\{L_k\}$ representing the coupling of the system to its environment via the \textit{master equation}
\begin{equation}\label{eq:master}
    \deriv{\rho(t)}{t} = \mathcal{L}[\rho(t)] := -i \left[H,\rho(t)\right] + \sum_k \left( L_k \rho(t) L_k^\dag - \frac{1}{2} \left\{ L_k^\dag L_k, \rho(t)\right\} \right)
\end{equation}
where $\{A,B\} = AB + BA$ is the anticommutator. Eq.~\eqref{eq:master} is the analog of a classical Langevin equation. The family of solution maps $\{\mathcal{E}_t\}_{t\ge 0}$, defined by $\mathcal{E}_t(\rho(0)) = \rho(t)$, forms a Markovian quantum dynamical semigroup, so that each $\mathcal{E}_t$ is completely positive and trace preserving. The dissipative term in the summation includes the action of a Kraus operator $L_k$ as in Eq.~\eqref{eq:ErhoC}, which by itself would increase the trace of $\rho(t)$. The anticommutator is included to correct this and keep the trace constant. Thus, the master equation is the most general continuous-time, Markovian, completely positive, trace-preserving evolution of a quantum state.

For example the discrete bit-flip channel with probability $p$ (see Eq.~\eqref{eq:Ebitflip}) can be represented in the continuous-time perspective with a single Lindblad operator $L = \sqrt{\gamma} X$. Here, $\gamma>0$ is the rate at which bits flip. In this case $L^\dag L = \gamma I$ and, in the absence of a Hamiltonian and for a single qubit, the master equation is
\[
\deriv{}{t}\rho(t) = \gamma\bigl(X\rho(t)X-\rho(t)\bigr).
\]
To solve it, decompose the initial state into components that are symmetric and antisymmetric under conjugation by $X$:
\[
\rho_m := \tfrac12\bigl(\rho(0)+X\rho(0)X\bigr),\qquad
\rho_d := \tfrac12\bigl(\rho(0)-X\rho(0)X\bigr),
\]
so that $\rho(0)=\rho_m+\rho_d$, with
\[
X\rho_m X-\rho_m=0,\qquad X\rho_d X-\rho_d=-2\rho_d.
\]
Using the ansatz $\rho(t)=\rho_m+\lambda(t)\rho_d$, substitution into the master equation gives
\[
\lambda'(t)=-2\gamma\,\lambda(t),\qquad \lambda(0)=1,
\]
hence $\lambda(t)=e^{-2\gamma t}$ and therefore
\begin{equation}
\rho(t)=\tfrac12\bigl(\rho(0)+X\rho(0)X\bigr)+\tfrac12e^{-2\gamma t}\bigl(\rho(0)-X\rho(0)X\bigr).
\end{equation}
Thus $\rho(t)\to \tfrac12\bigl(\rho(0)+X\rho(0)X\bigr)$ as $t\to\infty$, which is the expected stationary state of the continuous bit-flip dynamics.

\subsubsection{Non-Markovian noise}\label{sec:noise_model}

While the Lindblad equation captures \emph{Markovian} noise processes, experimental evidence on superconducting qubits, trapped ions, and spin qubits reveals significant \emph{non-Markovian} effects~\cite{white2020demonstration}. Non-Markovian noise arises when system dynamics are influenced by memory effects, meaning the past interactions with the environment affect its future evolution. Accurately modeling and mitigating non-Markovian noise is currently an active area of research, receiving much attention in recent studies, see, e.g., ~\cite{agarwal2024modelling,groszkowski2023,reddy2025,oda2024sparse} and references therein.

Recall, that mathematically Markovian dynamics are characterized by a time-local generator
\begin{equation}
    \deriv{\rc(t)}{t} = \mathcal{L}\rc(t),
\end{equation}
where $\mathcal{L}$ is a Lindbladian superoperator of the form Eq.~\eqref{eq:master} and $\rc$ is a computational state.  
In contrast, non-Markovian dynamics cannot, in general, be described by a time-local generator acting only on the instantaneous state. Instead, the system’s evolution depends explicitly on its history and is more naturally expressed through integro-differential equations with memory kernels.

Mathematically, the derivation can be obtained through the Nakajima-Zwan\-zig (NZ) projection operator formalism~\cite{nakajima1958quantum}, which is a part of the Mori-Zwanzig formalism~\cite{zwanzig2001nonequilibrium}. By starting from the unitary evolution of the joint system–environment state and eliminating the environmental degrees of freedom, the NZ approach yields an evolution equation for the reduced density operator corresponding to the ``relevant" degrees of freedom.

We outline the main idea of the NZ formalism and refer to, e.g.,~\cite{lidar2019lecture}, for more details. 
Recall, that for a fixed environment state $\re$, we can express the computational state $\rc =\tr_{E}(\rt)$, where $\rt$ is a total system-environment state. We define the projection superoperators
\begin{equation*}
    \pnz X =:\hat{X}= \tr_{E}(X)\otimes\re,\quad \qnz X =:\bar{X} = (I-\pnz)X.
\end{equation*}
The goal is to derive the evolution equation for $\hat{\rt} := \pnz\rt = \rc\otimes\re$ (the ``relevant'' degrees of freedom), corresponding to the computational state, by eliminating $\bar{\rt} := \qnz\rt$ (the ``irrelevant'' or environment degrees of freedom). Applying those operators to the total system-environment dynamics
\begin{equation}
    \deriv{\rt}{t} = \mathcal{L}_{CE
    }(t)\rt(t) = -i \left[{H}(t),\rt(t)\right],
\end{equation}
we can derive a coupled system of equations
\begin{align*}
    \deriv{\hat{\rt}}{t} = \hat{\mathcal{L}}_{CE}\hat{\rt} + \hat{\mathcal{L}}_{CE}\bar{\rt},\\
    \deriv{\bar{\rt}}{t} = \bar{\mathcal{L}}_{CE}\hat{\rt} + \bar{\mathcal{L}}_{CE}\bar{\rt},
\end{align*}
where $\hat{\mathcal{L}}_{CE}:=\pnz\mathcal{L}_{CE}$ and $\bar{\mathcal{L}}_{CE}:=\qnz\mathcal{L}_{CE}$.
Using variations of constants we can solve the second equation for $\bar{\rt}$ above and derive the corresponding NZ master equation for $\hat{\rt}$:
\begin{align*}
    \deriv{\hat{\rt}(t)}{t} = \hat{\mathcal{L}}_{CE}(t)\hat{\rt}(t) 
    + \hat{\mathcal{L}}_{CE}(t)\mathcal{G}(t,t_0)\bar{\rt}(t_0)
    +\int_{t_0}^t \hat{\mathcal{L}}_{CE}(t)\mathcal{G}(t,s)\bar{\mathcal{L}}_{CE}(s)\hat{\rt}(s)ds,
\end{align*}
where $\mathcal{G}(t,s)$ is the propagator of the irrelevant subspace.
The above equation is exact and makes explicit how non-Markovianity arises from persistent system-\-environment correlations. The terms on the right-hand side represent 
Markovian (memoryless part of the dynamics), inhomogeneous and the non-Markovian (memory part) terms, respectively.  We note that the homogeneous term, which depends on the initial condition, vanishes for a factorized initial condition, i.e., $\rt (0)= \rc(0)\otimes\re$.  

While NZ master equation is exact, it is not closed: the memory kernel is generally unknown and  encodes all effects of the eliminated environment, making its direct evaluation intractable. Consequently, one must either introduce analytical approximations or adopt data-driven approaches that infer the kernel directly from observed dynamics.

From a numerical standpoint, discretizing and simulating quantum master equations at scale presents both challenges and opportunities, such as balancing stability, accuracy, and structural properties.
Due to high-dimensionality direct time integration quickly becomes prohibitive for multi-qubit systems. To mitigate this, a range of structure-preserving approaches have been developed, including structure-preserving integrators, low-rank and tensor-network representations of density operators, Krylov and exponential integrator methods, and model reduction techniques that preserve complete positivity and trace, see, e.g.,~\cite{appelo2025kraus,le2015adaptive,lebris2013,cao2025structure,werner2016positive,weimer2021simulation} and references therein. These methods exploit the fact that physically relevant open quantum states often remain approximately low rank or weakly entangled over experimentally relevant time scales.

In the non-Markovian setting, memory terms introduce additional complexity beyond propagating a high-dimensional reduced state. Various numerical approaches are currently being investigated that employ compact kernel representations, such as finite-memory truncation, sum-of-exponentials (auxiliary-mode) approximations of memory kernels, and recasting non-Markovian dynamics into enlarged Markovian systems via auxiliary degrees of freedom; see, e.g.,~\cite{xu2026simulating,cerrillo2014,peng2025discretization,chen2020non,garraway1997nonperturbative,pollock2018non,strathearn2018efficient,luchnikov2020machine} and references therein.
Additionally, data-driven and learning-based methods are investigated to infer low-dimensional, interpretable memory representations directly from experimental or simulated data, either through kernel reconstruction or multi-time process characterization, enabling scalable and uncertainty-aware modeling of non-Markovian noise, e.g.,~\cite{luchnikov2020machine,white2022non}.

\subsection{Error correction and the code space} \label{sec:error_corr}

As mentioned in \sectionref{sec:Qchannels}, a \textit{quantum error} refers to the effective action of a noise channel on a quantum state that alters the intended computation. The reliable extraction of information from a quantum circuit thus becomes itself a problem of uncertainty quantification: we must detect errors, model them, and design ways in which they can be systematically avoided, mitigated and corrected. This is a large and very active field of research and we will limit ourselves to the basics, specifically to those aspects that resemble or require methods in uncertainty quantification. For a more thorough introduction we refer the reader to~\cite{roffe2019quantum, scherer2019mathematics}.

Detecting errors in quantum computing faces the challenge that measuring a quantum system  changes it irreversibly, which prevents us from applying a simple measurement, or repeated independent measurements, to determine to what extent has noise corrupted the state of the system. Moreover, the creation of  identical copies of quantum states is deemed impossible by the ``no-cloning theorem''. Therefore we cannot use simple comparisons between copies to detect the presence of errors. 

These challenges are addressed by introducing suitable redundancies in the way information is encoded in the states within a quantum circuit. This means expanding the Hilbert space in which the qubits are encoded and creating redundant \textit{physical qubits} to represent the target \textit{logical qubits}. For example, a pure state with a single logical qubit can be encoded in two physical qubits by the following encoder
\begin{equation}
    \ket{\psi} = \alpha \ket{0} + \beta \ket{1} \mapsto  
    \ket{\psi}_L = \alpha \ket{00} + \beta \ket{11} 
    = \alpha \ket{0}_L + \beta \ket{1}_L,
\end{equation}
where the $L$ subscript denotes the logical state version of the physical state. In this example we have expanded the Hilbert space from $\mathcal{H}_1$ to a subspace $\mathcal{C} = \text{span}\{\ket{00},\ket{11}\}$ of $\mathcal{H}_2$. The space $\mathcal{C}$ of logical qubits is called the \textit{code space}. The effect of this encoding is to distribute the information of $\ket{\psi}$ across the entangled two-qubit state $\ket{\psi}_L$. This encoding extends naturally to logical mixed states $\rho \mapsto \rho_L \in \mathcal{C}$.

Once a system with $n$ logical qubits is encoded in a code space $\mathcal{C} \subset \mathcal{H}_{n'}$ of $n' >n$ physical qubits, we must construct \textit{logical gates}  on $\mathcal{H}_{n'}$ that preserve the code space $\mathcal{C}$ and act as the original gates on it. Namely, if $U$ is an original gate on $\mathcal{H}_n$, its corresponding logical gate $U'$ on on $\mathcal{H}_{n'}$ must map $\mathcal{C}$ into itself and be such that $U' \ket{\psi}_L = \ket{U \psi}_L$ for any $\ket{\psi} \in \mathcal{H}_n$.

The quantum circuit of logical gates will be subjected to a noise channel $\mathcal{E}$ on $\mathcal{H}_{n'}$ as described in \sectionref{sec:env_noise}. A logical state $\rho_L$ is transformed by noise to $\mathcal{E}(\rho_L) = \sum_k E_k \rho_L E_k^\dag$, where $\{E_k\}$ is the Kraus decomposition of $\mathcal{E}$ and each $E_k$ can be thought as one possible detectable error type, with $E_0 = I$ being the possibility of no error. If the error subspaces $\{E_k \mathcal{C}\}$ are orthogonal, then one can perform a projective measurement $\{P_k\}$ that would output the \textit{syndrome} $k$ of the corresponding error, or 0 if there was no error. Further, this measurement will collapse the state to one proportional to $E_k \rho_L E_k^\dag$ thus identifying the error without disturbing the logical information. These type of measurements are called \textit{stabilizers}  in the context of error correction. Having detected the syndrome, error correction then introduces a \textit{recovery channel} $\mathcal{R}$ that undoes $E_k$ yielding $(\mathcal{R}\circ \mathcal{E})(\rho_L) = \rho_L$. For example, if $E_k$ is a unitary, then the corresponding Kraus operator of $\mathcal{R}$ is simply $R_k = E_k^\dag$. 

The possibility of detecting and correcting errors rests then in the construction of an encoding to an appropriate code space, the identification of a sufficiently rich noise model, and the implementation of the corresponding logical gates, stabilizer measurements and recovery maps. The Knill–Laflamme theorem gives the necessary and sufficient conditions for the existence of such recovery channel: for every $k,l$ there must exist some $c_{k,l} \in \mathbb{C}$ such that 
\begin{equation}
    P_{\mathcal{C}} E_k^\dag E_l P_{\mathcal{C}} = c_{kl} P_{\mathcal{C}} 
\end{equation}
where $P_{\mathcal{C}}$ is the projector on the code space. Namely, the error operator $E_k^\dag E_l$ must be proportional to the identity in $\mathcal{C}$. In this case, there will exist a recovery channel $\mathcal{R}$ such that $\mathcal{R} \circ \mathcal{E} = I$ on $\mathcal{C}$ and we would say that we have an \textit{error-correcting code}.

The framework of encoding, syndrome measurement, and recovery, forms the mathematical foundation of quantum error correction, ensuring that logical information can be protected against the uncertainty introduced by noise. While the theory is general, its most powerful realization to date is through surface codes, where logical qubits are encoded in two-dimensional lattices of physical qubits with local stabilizer checks. These codes exploit redundancy and locality to achieve remarkably high error thresholds, and have been implemented with great success in current experimental platforms. Surface codes like that reported by~\cite{google2023suppressing}  now stand as the leading architecture for fault-tolerant quantum computation, providing a practical route to controlling uncertainty in real quantum devices.


\section{Uncertainty Quantification Methods in Quantum Computing}
\label{sec:UQmethods}

Uncertainty quantification in quantum computing seeks to rigorously characterize, propagate, and reduce the various forms of uncertainty that arise in the preparation, evolution, and measurement of quantum states. Methods for UQ in this context therefore integrate mathematical tools from probability theory, functional analysis, and statistical estimation with the operator-theoretic framework of quantum mechanics. They include approaches for modeling the statistics of measurement outcomes (e.g., through Born’s rule and concentration inequalities), for estimating states and processes under noisy observations (e.g., quantum tomography and Bayesian inference), and for assessing algorithmic accuracy and stability under realistic error models (e.g., fidelity analysis, perturbative expansions, and master equation formalisms). Together, these methods provide the foundation for evaluating the reliability of quantum algorithms and for guiding the design of error mitigation and correction strategies, thereby linking the mathematics of uncertainty with the engineering of scalable quantum devices.

Our goal in this chapter is to represent some of the techniques of the literature beyond textbook. Namely, we organize methods and techniques according to the following graph. These are not mutually exclusive categories, but rather overlapping areas where UQ methods are applied in quantum computing. The methods themselves were selected based on our mathematical interests and expertise, and are not meant to be exhaustive. However, we wanted to be representative of the different areas where UQ is applied in quantum computing and provide enough technical details to be useful to the reader. To this end, we review some of the current practices in quantum computing that directly use, or closely parallel, established techniques in uncertainty quantification (UQ).

\noindent

 \resizebox{0.95\linewidth}{!}{\begin{tikzpicture}[
    every node/.style={},
    parent/.style={
        draw,
        rounded corners,
        thick,
        align=center,
        minimum width=2.5cm,
        minimum height=0.9cm
    },
    childbox/.style={
        draw,
        rounded corners,
        thick,
        inner sep=6pt,
        text width=4.5cm,     
        align=left,
        anchor=north
    }
]

\node[parent] (parent) at (0,0) {UQ for QC};

\def\colheight{8cm}  

\node[childbox] (child1) at (-5.5,-3.5) {%
  \begin{minipage}[t][\colheight][t]{\linewidth}%
    \centering
    \textbf{Sampling with}\\[2pt]
    \textbf{Quantum Algorithms}\\[4pt]
    \rule{\linewidth}{0.4pt}\\[6pt]
    \begin{enumerate}[leftmargin=*, itemsep=3pt, topsep=3pt]
      \item Measurement algorithm
      \item Probability estimation
      \item Amplitude estimation
      \item Optimal shot allocation
      \item Quasiprobabilities
      \item UQ in VQC
    \end{enumerate}
  \end{minipage}%
};

\node[childbox] (child2) at (0,-3.5) {%
  \begin{minipage}[t][\colheight][t]{\linewidth}%
    \centering
    \textbf{Resource}\\[2pt]
    \textbf{Characterization}\\[4pt]
    \rule{\linewidth}{0.4pt}\\[6pt]
    \begin{enumerate}[leftmargin=*, itemsep=3pt, topsep=3pt]
      \item Metrics \& benchmarking
      \item State characterization
      \item Process characterization and noise modeling
    \end{enumerate}
  \end{minipage}%
};

\node[childbox] (child3) at (5.5,-3.5) {%
  \begin{minipage}[t][\colheight][t]{\linewidth}%
    \centering
    \textbf{Error}\\[2pt]
    \textbf{Management}\\[4pt]
    \rule{\linewidth}{0.4pt}\\[6pt]
    \begin{enumerate}[leftmargin=*, itemsep=3pt, topsep=3pt]
      \item Sensitivity analysis
      \item Error mitigation
      \item Noise separation
      \item Error rate UQ
    \end{enumerate}
  \end{minipage}%
};

\draw[thick,-latex] (parent.south) -- ($(child1.north)+(0,0.15)$);
\draw[thick,-latex] (parent.south) -- ($(child2.north)+(0,0.15)$);
\draw[thick,-latex] (parent.south) -- ($(child3.north)+(0,0.15)$);

\end{tikzpicture}}%

In~\sectionref{sec:sampling}, we survey UQ methods used to determine how many times a quantum algorithm must be executed and measured in order to accomplish different computational tasks. Each such execution is called a shot, and shot estimation—viewed as the problem of balancing accuracy against computational resources—is a central point of contact between UQ and quantum computation. We show how UQ techniques yield shot-complexity bounds guaranteeing a desired precision and success probability for algorithms based on projective measurements. For expectation-value and probability-estimation algorithms, the familiar $1/\sqrt{N}$ Monte Carlo scaling serves as a baseline against which quantum improvements are measured. Building on this foundation, we discuss UQ-inspired quantum methods that coherently reshape probability distributions to achieve quadratic improvements in sampling cost and to enable high-precision estimation of success probabilities and expectation values. Some of these approaches, such as stratified sampling, are direct adaptations of classical UQ theory. Other approaches such as amplitude amplification and quasiprobability sampling, exploit intrinsically quantum structure to achieve resource savings unattainable classically. Together, these methods illustrate how uncertainty quantification provides both a conceptual framework and a practical toolkit for understanding and optimizing the sampling behavior of quantum algorithms. 

In~\sectionref{sec:characterization}, we examine resource characterization and testing through the lens of inverse problems: inferring quantum states, processes, and dynamical generators from finite, noisy, indirect measurements. Accurate characterization is foundational to quantum computing, yet the data available are inherently incomplete and corrupted by state-preparation-and-measurement (SPAM) errors, calibration drift, and stochastic fluctuations. Classical UQ techniques—ranging from maximum-likelihood estimation and Bayesian inference for quantum state tomography to bootstrapped confidence intervals in randomized benchmarking—enable practitioners to attach rigorous uncertainty bounds to reconstructed density matrices, gate fidelities, and Pauli error rates. We survey how metrics and benchmarking protocols compress noisy device behavior into interpretable parameters with explicit sampling-error estimates, how tomographic methods balance physical constraints against data fidelity, and how process characterization frameworks propagate calibration uncertainties through the reconstruction. These methods reveal that without principled uncertainty quantification, one cannot separate intrinsic device errors from reconstruction artifacts, making UQ central to trustworthy quantum device validation.

In~\sectionref{sec:error_management}, we address how UQ informs the design and assessment of protocols that manage errors in near-term quantum devices. We begin with sensitivity analysis, showing how variance-based methods such as parameter-shift gradient estimation quantify which gate parameters and noise sources most significantly affect algorithm output, thereby guiding targeted calibration and error-budget allocation. We then review error mitigation techniques—zero-noise extrapolation, probabilistic error cancellation, symmetry verification, and learning-assisted approaches—that systematically reduce bias despite introducing additional statistical uncertainty. In this context, rigorous UQ provides the tools to quantify and balance these effects, enabling transparent reporting of mean-squared error and informed decisions about when mitigation offers a net improvement over raw measurements. We also discuss importance-sampling strategies that efficiently estimate rare algorithmic failure events, and the hierarchical discrete fluctuation auto-segmentation (HDFA) framework that decomposes qubit noise into additive stochastic components on distinct timescales. Together, these tools exemplify how UQ bridges device characterization and algorithm execution, enabling practitioners to propagate hardware uncertainties into end-to-end performance guarantees and to maintain statistical coherence across the full quantum computing stack.

\subsection{Sampling with Quantum Algorithms}\label{sec:sampling}

In quantum algorithms, uncertainty quantification is embedded in the resource analysis: for each task, one must determine how many oracle queries or measurement shots (circuit repetitions) are required to guarantee that the output is within a given target accuracy with high probability. This is variously called sample complexity, query complexity, or accuracy–resource tradeoff analysis. One can broadly divide algorithms into those that require expectation value estimation, where the costs associated with Monte Carlo sampling and averaging determine the shot complexity, and those that rely on a single projective measurement of a suitably engineered register. The latter output an answer with quantifiable precision as a function of its computational resources.

\subsubsection{Projective-measurement algorithms}

We refer here to algorithms that use carefully designed projective measurements to extract information from a suitably engineered register in the output of a quantum circuit. This includes paradigmatic quantum algorithms such as quantum phase estimation and Grover's search algorithm. In these cases, uncertainty quantification considers the resources needed to obtain the desired measurement outcome with high enough probability.

A typical example of a projective-measurement algorithm is \textit{phase estimation}. This algorithm is a common subroutine in many useful quantum computations, from ground-state energy calculations in quantum chemistry, to Shor's factorization algorithm. The goal is, given a unitary $U$ on $\mathcal{H}_n$, to estimate the phase $\theta$ of an eigenvalue $e^{2 \pi i \theta}$ of $U$. The algorithm uses $n'$ additional \textit{ancilla} qubits, so it operates on $\mathcal{H}_{n+n'}$. The output of the algorithm is a computational basis measurement on the ancilla register of the output state $\rho_{\tout} \in \mathcal{H}_{n+n'}$ (see Eq.~\eqref{eq:probMx}) providing a bitstring $k_1,\dots,k_{n'}$ such that $\hat{\theta} = 0.k_1k_2\dots k_{n'}$ approximates $\theta$. The output is probabilistic even in the absence of noise; in this case the probability of error has upper bound 
\begin{equation}\label{eq:acc_phase}
    \P_{\rho_{\tout}}\left(|\theta - \hat{\theta}| < \epsilon\right) > 1-\frac{1}{2(2^{n'} \epsilon + 1)}
\end{equation}
from which estimation of the required number of ancilla qubits and number of shots required for a given accuracy and likelihood can be derived, see, e.g.,~\cite[Section~2.5]{nielsen2010quantum}.

\subsubsection{Expectation value and probability estimation}\label{sec:ExpProbEst}

Estimating the expectation value of a quantum circuit output observable to a given accuracy generally requires repeating the entire quantum circuit (from input to final measurement) many times. These are referred as \textit{shots}, with each shot producing a probabilistic outcome (an eigenvalue of the observable, distributed according to Born's rule in Eq.~\eqref{eq:obsm}), so the observable’s expectation value must be inferred statistically from the sample. Classical statistical estimation techniques apply directly to this process: the sample mean of $N$ independent measurement outcomes is an unbiased estimator of the true expectation, and its uncertainty decreases as $N$ increases. In fact, the standard error (standard deviation of the sample mean) typically scales as $\propto 1/\sqrt{N}$ – often called the “shot noise” limit – the same scaling familiar from classical Monte Carlo integration and the law of large numbers.

To be precise, consider a quantum circuit operator $\rho = U \rho_{\tin} U^\dag$ in $\mathcal{H}_n$, and an observable $O$ with spectrum $\{o\}$, for which we want to compute $\EXP_{\rho}(O)$. Assuming that for each shot the state $\rho_{\tin}$ is perfectly prepared and the unitary $U$ is identical, then $N$ shots will produce a sequence of random variables $X_1,\dots,X_N$ on $\{o\}$ each distributed according to $\P_{\rho}(\text{observe } o) = \tr(P_o \rho)$. The law of large numbers applies directly to this case: the mean $\bar{X}_N = \frac{1}{N} \sum_j X_j$ is an unbiased estimator for $\EXP_{\rho}(O)$ and the mean square error is 
\begin{equation}\label{eq:MSE}
    \EXP \left|\bar{X}_N - \EXP_{\rho}(O) \right|^2 = \frac{1}{N} \Var_{\rho}(O)
\end{equation}
where $ \Var_{\rho}(O)$ denotes the quantum variance of $O$ in $\rho$. Note the use of the subscript $\rho$ to distinguish expectation in quantum probabilities from sample means. The quantum variance can be written simply as the variance of a random variable taking values on the spectrum $\{o\}$ of $O$ with probabilities $\tr(P_o \rho)$:
\begin{equation}\label{eq:qVar}
    \Var_{\rho}(O) = \tr(O^2 \rho) - \tr(O\rho)^2 = \sum_o \tr(P_o \rho) o^2 - \left(\sum_o \tr(P_o \rho) o \right)^2 \leq \frac{1}{4}(o_{\max} - o_{\min})^2
\end{equation}
where $o_{\max}$ and $o_{\min}$ are the maximum and minimum eigenvalues of $O$ respectively and the last bound follows from Popoviciu's inequality.

A simple concentration bound can be obtained from Hoeffding’s inequality for any accuracy $\epsilon > 0$
\begin{equation}
    \P\left(\left| \bar{X}_N - \EXP_{\rho}(O) \right| \geq \epsilon\right) \leq 2 \exp\left(-\frac{2 N \epsilon^2}{(o_{\max} - o_{\min})^2}\right).
\end{equation}
It follows that the number of shots $N$ required to obtain a measurement of accuracy $\epsilon$ with probability larger than $1-\delta$ is at least
\begin{equation}\label{eq:Nshots}
    N(\epsilon,\delta) \geq (o_{\max} - o_{\min})^2 \frac{\log(2/\delta)}{2 \epsilon^2}.
\end{equation}
Namely, it scales polynomially in $1/\epsilon$. Estimators of this kind are called ``polyprecision estimators'', in contrast with exponential-precision estimators that scale logarithmically in $1/\epsilon$.

A related problem is that of estimating the quantum probability $\P_{\rho}(\text{observe } o=(o_1,\dots,o_n))$ where $o_j$ denotes the outcome of measuring of the $j$ qubit on the output state of the quantum circuit. This probability can be estimated by performing $N$ shots of the circuit and measuring the frequency $f_N(o)$ of occurrence of each possible vector $o$. In this case the spectrum of each measurement is bounded by one and Eq.~\eqref{eq:Nshots} gives an polyprecision bound $N(\epsilon,\delta) \geq \log(2/\delta)/2 \epsilon^2$ for the number of shots required to measure $f_N(o)$ within an accuracy $\epsilon$ of the target with probability $1-\delta$ ~\cite{pashayan2015estimating}.

\subsubsection{Amplitude amplification and estimation}

The aim of many quantum computations, e.g. search algorithms, is to return a single ``good'' basis state $\ket{x^*}$ from a superposition $\ket{\psi} = \sum_x \alpha_x \ket{x}$. For example, Grover's search algorithm encodes all possible values of a system in the input state, then applies a \textit{quantum oracle} that marks the desired state $\ket{x^*}$. Let $p^* = |\alpha_{x^*}|^2$. Naively, an average of $O(1/p^*)$ shots are needed to observe the target state. Grover's algorithm invokes the oracle $O(1/\sqrt{p^*})$ times prior to measurement in such a way that the probability of observing $x^*$ in the final measurement is close to one,~\cite[Chapter 6]{nielsen2010quantum}.

Grover's search algorithm is the canonical example of \textit{amplitude amplification}. To illustrate it, consider the state $\ket{\psi}$ decomposed it into its target and the rest, 
\begin{equation*}
    \ket{\psi} = \alpha_{x^*} \ket{x^*} +  \sum_{x \neq x^*} \alpha_x \ket{x}.
\end{equation*}
Write the target coefficient in terms of its amplitude and phase as $\alpha_{x^*} = \sqrt{p^*} e^{i \phi^*} $. If we further write $\sqrt{p^*} = \sin(\theta^*)$ for some angle $\theta^*$, then the full state is $\ket{\psi} = \sin(\theta^*) e^{i \phi^*} \ket{x^*}+ \cos(\theta^*) \ket{\tilde{\psi}}$ for some state $\ket{\tilde{\psi}}$ orthogonal to $\ket{x^*}$.  The quantum oracle $O$ marks $\ket{x^*}$  by flipping its phase, namely 
\begin{equation*}
   O \ket{\psi}  = -\sin(\theta^*) e^{i \phi^*} \ket{x^*}+ \cos(\theta^*) \ket{\tilde{\psi}}
\end{equation*}
Now, consider a \textit{diffusion operator} $D$ given by the reflection across $\ket{\psi}$, $D = 2 \ket{\psi}\bra{\psi} - I$. Grover's iteration, the amplitude amplification operator in Grover's algorithm is, $G = DO$. It can be shown that $G$ is equivalent to a rotation by $2 \theta^*$ and that $k$ applications of $G$ give the state 
\begin{equation}\label{eq:Grovers}
    G^k \ket{\psi} = \sin((2k+1) \theta^*) e^{i \phi^*} \ket{x^*} + \cos((2k+1) \theta^*) \ket{\tilde{\psi}}.
\end{equation}
If the initial amplitude $\sqrt{p^*}$ is very small, so is the initial angle $\theta^*$. By rotating the state towards $\ket{x^*}$, Grover's operator can make $(2k+1)\theta^* \approx \pi/2$ in $k\approx \frac{\pi}{4 \theta^*} - \frac{1}{2} = O(1/\sqrt{p^*})$ iterations.

The key insight behind these algorithms, as introduced in~\cite{brassard2000quantum}, is that quantum operations act on amplitudes, not probabilities. Specifically, one can perform a sequence of rotations that increases the target amplitude $\sqrt{p^*}$ by a constant factor per iteration, where $p^* = |\alpha_{x^*}|^2$ is the initial success probability. After $k = O(1/\sqrt{p^*})$ such iterations, the resulting state yields $x^*$ with high probability in a single measurement. The success probability after $k$ iterations is given by $p^{(k)} = \sin^2\big((2k+1)\theta^*\big)$. In general, amplitude amplification reduces uncertainty by shifting quantum computational resources from $O(1/p^*)$ repeated measurements to $O(1/\sqrt{p^*})$ coherent operations.

A comprehensive treatment of amplitude amplification can be found in~\cite{kwon2021quantum}. It can be used in general to amplify the amplitude of the target component of a quantum state (even if this target has dimension larger than one). Besides quantum search algorithms, amplitude amplifications is used in algorithms for quantum mean estimation~\cite{shyamsundar2023non}, optimization~\cite{matwiejew2023quantum}, and linear algebra~\cite{ambainis2012variable}. 

While amplitude amplification boosts the probability of observing a desired outcome by coherently rotating the state into the good subspace, \textit{amplitude estimation} combines this operator with quantum phase estimation to determine the success probability $p^*$ of the state $\ket{\psi}$. It is, in a sense, an uncertainty quantification method. These methods are central in quantum algorithms for expectation estimation, quantum Monte Carlo, counting, optimization, and a variety of statistical and machine-learning tasks where estimating probabilities or averages is the primary computational goal, see~\cite{brassard2000quantum}.

Additionally, note that in order to apply Grover rotations in Eq.~\eqref{eq:Grovers} an optimal number of times requires knowing the angle $\theta^*$. In practice, however, the target probability $p^*$ is usually unknown. Although there exist robust versions of Grover's algorithm that do not require prior knowledge of this amplitude, amplitude estimation can be used in addition to amplitude amplification for optimal implementation of Grover's search. Specifically, since Grover's operator acts as a rotation by angle $2 \theta^*$, it can be shown that its eigenvalues are $e^{\pm 2 i \theta^*}$. A phase estimation with $m$ ancilla qubits, uses controlled powers $G^j$, $j=0,\dots,2^{m}-1$ to coherently estimate the eigenphase $\theta^*$ and hence $p^*$ with an additive error of $O(1/2^m)$, see~\cite[Chapter 5]{nielsen2010quantum}. 

Modern variants of amplitude estimation seek to address the common challenges such as the need for deep circuits, large ancilla registers, and controlled applications of large powers of $G$ that are costly to implement. They also incorporate ideas from classical statistical estimation theory.  Such methods include the iterative amplitude estimation~\cite{suzuki2020amplitude}, the maximum likelihood method of  ~\cite{grinko2021iterative} and the robust amplitude estimation procedure of~\cite{tanaka2022noisy}.

\subsubsection{Optimal shot allocation}

When multiple observables must be estimated simultaneously,  the central question becomes how to distribute a limited budget of measurements in order to estimate the the different terms. This problem is known as optimal shot allocation, and it plays a role in quantum uncertainty quantification that is directly analogous to stratified sampling in classical statistics. By allocating more shots to observables with larger coefficients and higher intrinsic quantum variance, one can minimize the overall estimator variance and thus improve the precision of quantum expectation values.

To be specific, suppose that we wish to compute the expectation $\EXP_\rho H = \tr(H \rho)$ of a Hamiltonian $H$ acting on $\mathcal{H}_n$, where $H$ can be decomposed into a weighted sum of simpler Hermitian observables $H = \sum_i c_i H_i$. This situation arises, for example, in the Variational Quantum Eigensolver technique used for solving the electronic structure problem of chemical physics. In this case the electronic Hamiltonian of a  molecular configuration is expressed as a linear combination of Pauli strings $H_i \in \{I,X,Y,Z\}^{\otimes n}$, see, e.g.,~\cite{verteletskyi2020measurement}. Each $H_i$ can be measured on its own basis using $N_i$ shots to obtain an estimate $\hat{H}_{i,N_i}$ of $\EXP_\rho H_i$. The total number of shots is $N =\sum_i N_i$ and the full estimate is $\hat{H}_N = \sum_i c_i \hat{H}_{i,N_i}$. The optimal allocation problem is then
\begin{equation}\label{eq:optimal_shot}
    \min_{\{N_i\}} \sum_i \frac{c_i}{N_i} \Var_{\rho}(H_i), \text{  subject to  } \sum_i N_i = N
\end{equation}

The solution to Eq.~\eqref{eq:optimal_shot} is the analog of the Neyman allocation for stratified sampling. Refinements to this techniques include grouping commuting operators~\cite{anschuetz2019variational}, using the spectrum of $H$ to decide optimum allocation~\cite{knill2007optimal}, or distributing shots among multiple quantum processors~\cite{bisicchia2024distributing}

\subsubsection{Using quasiprobabilities}

A quasiprobability representation of the type Eq.~\eqref{eq:totalQP} provides an interpretation of the Born rule as the expectation value of a stochastic process~\cite{pashayan2015estimating}. Note that if all of the terms within each summand of Eq.~\eqref{eq:totalQP} is a true probability, then the quantum expectation could be simulated via Monte Carlo by sampling $\lambda$, evaluating a Markov transition $\lambda \to \lambda'$ and evaluating $O$. Quasiprobaility sampling approaches design classical Markov chains $\{\lambda_k\}_{k=0}^{L}$ and appropriate random variables $X_k$ such that
\begin{equation}\label{eq:EXPquasi}
    \EXP_{\rho_\tout}(O) = \EXP(X_L(\lambda_L) \dots X_0(\lambda_0)).
\end{equation}
This is accomplished by carefully moving the negative signs of the quasiprobability distributions in Eq.~\eqref{eq:totalQP} to the variables $X_k$. 

The amount of ``negativity'' in the quasiprobabilities  distribution of $\rho$ is $N(\rho) = \sum_{\lambda} |\tr(G_\lambda \rho)|$. This negativity is used for bounding the variance of the estimator in Eq.~\eqref{eq:EXPquasi}. Under certain regimes, the resulting sampling can provide polyprecision estimators with better performance than that of Eq.~\eqref{eq:Nshots} for the direct Monte Carlo outlined, see, e.g.,~\cite{pashayan2015estimating}. 

Notably, simulation via quasiprobability representations transforms the problem of simulating quantum mechanics into a statistical estimation problem, for which uncertainty quantification methods (variance reduction, importance sampling, concentration bounds) can be applied. 

\subsubsection{UQ in variational quantum circuits}

Recall from \sectionref{sec:VQC} that variational quantum circuits are used as efficient evaluators of a cost function\linebreak[0] $C(\theta) = \braket{\psi(\theta) | O| \psi(\theta)}$, where $\ket{\psi(\theta)} = U(\theta) \ket{\psi_0}$ is the parametrized output of the circuit and $O$ is some observable. Moreover, the circuit can be used to evaluate the gradient $\partial C(\theta)/\partial \theta$ through the parameter-shift rule in Eq.~\eqref{eq:paramShift}. This information is passed to a classical optimizer that produces updates on $\theta$ until convergence to a minimizer $\theta^*$ of $C(\theta)$.

Uncertainty permeates every step of this process. For each fixed value of $\theta$, the cost function is, in reality, the quantum expectation $C(\theta) = \EXP_{\ket{\psi(\theta)}}(O)$, and its numerical evaluation is subject to the statistical uncertainties discussed in \sectionref{sec:ExpProbEst}. Even in the absence of hardware noise, a hybrid variational algorithm accesses only the shot-based estimator $\overline{C(\theta)}_N$, computed from $N$ projective measurements. Thus the algorithm effectively samples a random field $\theta \mapsto \overline{C(\theta)}_N$ rather than the true cost landscape. By Eq.~\eqref{eq:MSE} these estimators introduce statistical variances,
\begin{equation}
    \Var [\overline{C(\theta)}_N] = \text{O}\left(\frac{1}{N}\right), \quad 
    \Var \left[\left(\overline{
    \frac{\partial C(\theta)}{\partial \theta}}\right)_N\right] = \text{O}\left(\frac{1}{N}\right)
\end{equation}
so that both the cost and its gradient are inherently noisy quantities. As a result, the variational procedure is, in practice, a stochastic gradient descent algorithm, whose performance depends critically on how measurement resources are allocated. A detailed discussion of adaptive shot-allocation strategies ensuring reliable convergence of variational quantum circuits can be found in~\cite{kubler2020adaptive}.

An alternative approach to stochastic gradient descent methods is to build a surrogate stochastic model for the cost function $C(\theta)$ from the noisy observations. Finding the optimal parameters $\theta^*$, thus becomes a Bayesian optimization problem~\cite{self2021variational}. Gaussian processes can serve as effective surrogate models for variational quantum algorithms. In particular, one can design kernel functions tailored to the specific circuit unitary~\cite{smith2023faster}, achieving higher-accuracy solutions with significantly fewer measurement shots. 

Another area of research related to uncertainty quantification in the context of variational quantum circuits relates to their \textit{expressiveness}. This concept, introduced in~\cite{sim2019expressibility}, broadly refers to the size and richness of the family of quantum states $\ket{\psi(\theta)}$ an ansatz can generate when $\theta$ varies uniformely over its range in $\mathbb{R}^p$. 

Mathematically, expressiveness is quantified by considering the probability distribution over unitaries induced by sampling parameters $\theta$ uniformly from $[-\pi,\pi]^p$ and mapping them via $U_\#: \theta \mapsto U(\theta)$. One then measures how closely this induced distribution approximates the Haar measure, i.e., the uniform distribution over unitary operators.
Various measures of expressiveness have been proposed, including some based on the Kullback-Leibner divergence as in~\cite{hubregtsen2021evaluation}, and the covering number in~\cite{du2022efficient}. In the original paper by~\cite{sim2019expressibility} the authors propose that $U_\#$ should approximate a ``unitary $t$-design'', namely that 
\begin{equation}
    \EXP_{U \sim U_\#}\left[ U^{\otimes t}\otimes U^{\otimes t}\right] \approx \EXP_{U \sim \text{Haar}}\left[ U^{\otimes t}\otimes U^{\otimes t}\right]
\end{equation}
which means that the $t$-moment of $U_\#$, also known as the  `frame potential',  is approximately equal to the corresponding moment with Haar measure. By choosing a $t$-design ansatz, one ensures a variational circuit that efficiently samples uniformly the space of unitaries and has thus high probability of generating a state that is close to the target. For ansatz $U(\theta)$ that have commutative generators,~\cite{ramirez2024expressiveness} show that the frame potential can be computed using uncertainty quantification techniques to sample an appropriate random walk. 

A fundamental challenge in the scalability of variational quantum circuits is the phenomenon of \textit{barren plateaus}, first identified by~\cite{mcclean2018barren}. A barren plateau is a large region of the parameter space in which the cost function landscape becomes exponentially flat, meaning that the gradient magnitude satisfies $|\partial C/\partial \theta| = \text{O}(2^{-n})$ where $n$ is the number of qubits. It has been understood that this geometrical feature of the cost emerges as expressiveness increases, because the induced distribution of states and gradients inherits the concentration-of-measure phenomena of Haar-random unitaries. In this regime, gradients vanish exponentially and become overwhelmed by sampling noise, making variational optimization statistically infeasible, see~\cite{cunningham2024investigating} for a comprehensive recent review on the causes and mitigation techniques.

\subsection{Resource characterization and testing}
\label{sec:characterization}

Accurate characterization of quantum resources, such as quantum states, quantum processes, and their dynamical generators, can be viewed fundamentally as an inverse problems: one observes measurement outcomes generated by an unknown quantum object (a state, a process, or a dynamical model) and seeks to infer that object from finite, noisy, indirect data.

For example, in quantum state tomography, the goal is to reconstruct a density operator that best explains a set of measurement outcomes; in quantum process tomography and gate-set tomography, the task extends to estimating unknown quantum channels and the gates that compose a computational architecture. These reconstruction problems are inherently ill-posed: the available data are finite, noisy, and incomplete. As a result, even in idealized settings, multiple states or processes may be consistent with the same observations, making careful modeling of uncertainty indispensable.
Reliable resource characterization must therefore account not only for reconstructing states and processes but also for quantifying the uncertainty inherent in those reconstructions. 
Modern quantum processors exhibit stochastic fluctuations, calibration drift, and non-negligible SPAM errors, all of which introduce systematic and statistical uncertainty into the reconstructed objects. Standard tomographic methods often neglect these effects and can yield biased, over-confident estimates; for example,~\cite{merkel2013} showed that substantial discrepancies arise when SPAM error is not explicitly modeled. Without an assessment of uncertainty, it becomes impossible to separate intrinsic device errors from artifacts of the reconstruction procedure. Thus, UQ is central to modern quantum characterization: it supplies confidence regions and error estimates that make tomographic reconstructions interpretable, supports principled model selection, and guides both experiment design and device benchmarking.

In this section we provide an overview of several characterization methods and highlight the role and importance of UQ for trustworthy quantum device characterization. For more details on characterization and benchmarking of quantum devices we refer to, e.g.,~\cite{hashim2025,nielsen2010quantum,eisert2020quantum} and references therein.

\subsubsection{Metrics and Benchmarking}

Benchmarking protocols compress noisy device behavior into a few interpretable quantities while explicitly tracking SPAM effects and sampling error. Mathematically, each protocol constructs an estimator for a parameter of a noisy quantum channel $\mathcal{E}: \mathcal{H}_n \to \mathcal{H}_n$ (or a circuit of such channels) and supplies uncertainty bounds derived from the empirical frequencies of measurement outcomes in the computational basis $\mathcal{B}_n$. The underlying random variables are shot-level outcomes $\{0,1\}^n$ with probabilities governed by Born’s rule.

A practical taxonomy ties UQ to three layers of metrics~\cite{hashim2025}: (i) gate- or cycle-level error rates parameterizing stochastic and coherent deviations of channels $\mathcal{E}$ from ideals $\mathcal{U}$; (ii) circuit-level metrics (volumetric or mirror benchmarking) that map circuit width $w$ and depth $d$ to a success probability $\pi(w,d)$; and (iii) application- or logical-level metrics, such as logical error per cycle in a code patch, that forecast end-to-end algorithm success. Credible bounds on primitive parameters (e.g., depolarizing probability $\lambda$ or Pauli error rates) propagate through these layers to confidence intervals on $\pi(w,d)$ and on logical failure probabilities.

\textbf{Randomized benchmarking} (RB) samples \emph{Clifford words} of length $m$—that is, sequences of $m$ gates drawn from the Clifford group, a set of efficiently implementable unitaries that map Pauli operators to Pauli operators under conjugation—applies their inverses, and measures a survival probability $s_m$, the empirical frequency that the final measurement matches the prepared computational basis state. Under a depolarizing channel model, the expectation obeys $\mathbb{E}[s_m]=A p^{m}+B$ with $p\in(0,1]$ related to the average gate infidelity. UQ enters via (i) binomial variance of $s_m$ from finite shots, (ii) regression uncertainty in estimates $(\hat p, \hat A, \hat B)$, and (iii) model-selection tests that detect drift or gate-dependence when residuals deviate from the single-exponential model. Credible intervals for the average error rate are obtained by bootstrap over random sequences and shots or by Bayesian linear regression with prior on $p$~\cite{hashim2025}.

\textbf{Cycle benchmarking} replaces Clifford words with device-native cycles $C$ (parallel layers of gates) and reports Pauli error rates by fitting decay of measured Pauli expectation values $\braket{P}$ under repeated application of $C$. If $\Lambda_C$ is the Pauli transfer matrix of the noisy cycle, its eigenvalues $\{\lambda_j\}$ determine the decay $\mathbb{E}[\braket{P_m}] \approx \lambda_j^{m}$ for a Pauli $P$ in the $j$-th eigenspace. Linear fits of $\log \braket{P_m}$ yield estimators $\hat \lambda_j$ with uncertainties from shot noise; bootstrapping across seeds and Paulis produces confidence regions for each $\lambda_j$. \textbf{Mirror and volumetric benchmarking} similarly sample circuits of width $w$ and depth $d$ and estimate a success probability surface $\pi(w,d)$; confidence bands in $(w,d)$ reveal layout fragility and calibration drift~\cite{hashim2025}.

\textbf{``Quantumness'' metrics.} Entanglement witnesses and magic monotones are expectation values of Hermitian operators $W$ (or nonlinear functionals) that breaks some constraint that all classical states must satisfy, thereby certifying non-classical behavior (such as contextuality). Given a state $\rho$, one measures $\braket{W}=\tr(W\rho)$ from repeated shots, obtaining an estimator $\widehat{\braket{W}}$ with variance determined by the spectrum of $W$ and shot count. Confidence intervals that exclude the classical bound certify entanglement or contextuality. Cross-entropy benchmarking (XEB) estimates the fidelity of a random circuit to an ideal unitary $U$ by comparing measured probabilities $p_{\text{exp}}(x)$ to ideal $p_{\text{ideal}}(x)=|\braket{x|U\ket{0^{\otimes n}}}|^{2}$ through an empirical cross-entropy; variance across instances and shots quantifies confidence in XEB scores~\cite{eisert2020quantum}.

\textbf{Variational and application-focused benchmarks.} For variational circuits with parameters $\theta$, one estimates an energy $E(\theta)=\tr(\rho_{\theta} H)$ for some observable $H$ via sample means of Pauli terms. The estimator $\hat E$ aggregates independent Bernoulli outcomes for each term; its variance combines shot noise, optimizer randomness in $\theta$, and drift in the implemented channel. The robustness of results can be effectively communicated by reporting medians (or trimmed means) over random seeds with bootstrap confidence intervals. Gradients computed via parameter-shift rules are likewise empirical averages of differences of expectation values and inherit similar variance structure, so their confidence intervals indicate optimizer reliability.

Collectively, these protocols convert noisy quantum channels on $\mathcal{H}_n$ into statistical estimators whose distributions are controlled by shot counts, SPAM parameters, and drift. The resulting error bars allow benchmarking claims to be stated as bounds on algorithm success probabilities and fault-tolerance budgets, and they set priors for the subsequent state and process characterization tasks.

\subsubsection{State characterization}

``A basic yet important step in quantum information processing is the efficient and reliable characterization of quantum states~\cite{yu2022statistical}.'' State characterization involves defining and measuring the properties of quantum states. The state of a qubit can be represented as a vector in a two-dimensional complex Hilbert space, while for composite systems, the state space is typically constructed using the tensor product. One type of characterization may involve distinguishing between separable states and entangled states. The density operator (or density matrix) can be used to describe ensembles or mixed states, and the Schmidt decomposition to characterize entanglement.

\textbf{Quantum State Tomography (QST)} is an experimental procedure used to determine an unknown quantum state by working with its density matrix. The state of a $d$-qubit quantum device is generally described by a density matrix $\rho$. To fully specify an arbitrary quantum state, $2^{2d} - 1$ real parameters must be characterized. A single measurement of an observable yields information about at most $2^d - 1$ parameters, meaning a single measurement is not enough to determine the quantum state. Thus, repeated measurements of at least $2^{d}+1$ mutually unbiased bases are necessary to estimate every parameter of the state.  This requires repeated preparations and measurements, in different ways, of the same state to build up a complete description. QST's usefulness lies in providing a complete, model-independent characterization of a quantum state, enabling diagnosis, validation, and benchmarking of quantum devices. Once the density matrix is reconstructed, one can compute derived quantities such as fidelities, distances to target states, or entanglement measures. However, because full state reconstruction requires resources that scale exponentially with system size, QST suffers from the curse of dimensionality and is often impractical for large systems. Furthermore, the estimated density matrix is often influenced by SPAM errors, which must be carefully accounted for in the QST process to obtain an accurate representation in the first place. 

More specifically, tomographic measurements are implemented through a tomographically complete measurement scheme, which can be realized either as a \emph{single} informationally complete POVM or as a \emph{collection} of POVMs (e.g., projective measurements in different bases). In practice, a collection may be experimentally preferred even though one POVM is sufficient to perform the full tomography. This is because hardware often naturally measures in one basis at a time. Implementing multiple settings may be easier to calibrate and reduce comlexity over a collection. Mathematically, we can denote the measurement effects by $\{M_k^{(m)}\}_{k=1}^{2^d}$ for settings $m=1,\dots,M$. For an unknown state $\rho$, Born’s rule gives
\[
p_k^{(m)}=\tr\!\left(M_k^{(m)}\rho\right).
\]
If setting $m$ is repeated $N_m$ times and outcome $k$ is observed $n_k^{(m)}$ times, the empirical frequencies are
\[
\hat p_k^{(m)}=\frac{n_k^{(m)}}{N_m},\qquad \sum_{k=1}^{2^d} n_k^{(m)}=N_m.
\]
Tomographic completeness means that these probabilities provide enough independent information to reconstruct $\rho$ (up to statistical uncertainty). A common reconstruction strategy is to expand the state in an operator basis \(\{B_j\}_{j=1}^{2^{2d}-1}\) (e.g., tensor products of Pauli operators),
\[
\rho=\frac{I}{2^d}+\sum_{j=1}^{2^{2d}-1} r_j B_j.
\]
The tomographic estimates \(\hat p_k^{(m)}\) then define a linear inverse problem for the coefficients \(\{r_j\}\). Although linear inversion is simple and direct, finite-sample noise can produce nonphysical estimates (for example, matrices that are not positive semidefinite). To enforce physicality, tomography is often formulated as a constrained optimization or statistical inference problem. A standard choice is maximum-likelihood estimation (MLE), which finds the density matrix maximizing the likelihood of the observed counts under Born’s rule. MLE returns valid quantum states but can introduce bias, and uncertainty quantification around the estimate requires additional tools. Bayesian approaches treat \(\rho\) as a random variable and update a prior with measurement data to obtain a posterior over states, naturally yielding credible regions at higher computational cost.

More recent methods, such as compressed-sensing tomography, shadow tomography, and neural-network-assisted estimators (some of which are described below) aim to reduce the exponential sampling burden by exploiting structural assumptions (e.g., low rank). These techniques again rely on UQ to determine when the assumptions are valid, how reliable the reconstructions are, and how uncertainty propagates to downstream tasks such as fidelity estimation and benchmarking.

\textbf{Shadow tomography} moves away from the goal of full state reconstruction by focusing instead on predicting specific properties of the quantum system, which are often linear functions of the underlying density matrix $\rho$, such as the expectation values of a set of observables. This method involves creating a ``classical shadow'' ($S_{\rho}$), which is an approximate classical representation of the quantum state, constructed from randomized measurements. The process requires repeatedly applying a random unitary transformation (e.g., from an ensemble of unitaries like Clifford circuits or tensor products of single-qubit Clifford circuits) and then measuring all qubits in the computational basis. A key advantage of classical shadows is that the number of measurements ($N$) required to accurately predict $M$ different target functions scales logarithmically with $M$. This measurement count is independent of the actual system size (number of qubits, $n$) for many important properties, which avoids the exponential scaling of full tomography. Shadow tomography allows for efficient prediction of linear function values (such as fidelity or local observables) using a simple median-of-means protocol in the classical post-processing phase. Classical shadows are particularly useful for estimating expectation values of many local observables, subsystem entanglement entropies, and two-point correlation functions. This protocol is asymptotically optimal for prediction tasks, matching information-theoretic lower bounds. Furthermore, it has the advantage of enabling the selection of target properties to predict after the measurements have been completed~\cite{huang2020predicting}.

\textbf{Tensor Network State (TNS)} methods approximate the quantum state as a network of rank-$k$ tensors, where the bond dimension ($\chi$) determines the amount of entanglement that can be represented.  Typically, the Schmidt decomposition is used to truncate the representation to the $\chi$ most strongly entangled modes at each network bond. The cost of computing a TNS scales polynomially with $\chi$, but highly entangled systems can require exponentially large values of $\chi$ to be represented accurately~\cite{kim2023evidence}. Nevertheless, TNS representation can be quite expedient in many cases.  One example is matrix product state tomography, which requires only a polynomial number of samples to fully characterize a state provided it has a compatible entanglement structure~\cite{guo2024quantum}.

\subsubsection{Process characterization and noise modeling}
\label{sec:process_characterization}

Process identification refers to the task of modeling and estimating the dynamical rules that govern the evolution of quantum systems, with a focus on characterizing how quantum operations deviate from their ideal behavior by explicitly learning noise models from experimental data. Unlike state identification, which reconstructs static quantum states, process identification captures the full input–output behavior of gates, channels, or continuous-time dynamics. For example, this includes \emph{process tomography}, which reconstructs the action of a single gate, and \emph{gate set tomography}, which provides a self-consistent description of an entire instruction set. More physically grounded approaches involve \emph{Hamiltonian learning}, where the unitary part of the dynamics is inferred, relevant to closed quantum systems as discussed in \sectionref{sec:Qchannels}. Extension of this framework to open systems involves Lindblad master equation parameter estimation, by learning both coherent Hamiltonian contributions and dissipative noise processes.

Mathematically, process characterization is also an inverse problem of estimating model parameters from data. This can be realized using various techniques, ranging from constrained minimization problem to probabilistic approaches, such as Bayesian inference.  

\textbf{Quantum Process Tomography (QPT)} aims to reconstruct an unknown quantum operator $\mathcal{E}$, a CPTP map $\mathcal{E}\colon \rho \to\mathcal{E}(\rho)$ as introduced in \sectionref{sec:Qchannels}, from input-output measurement data.
Unlike quantum state tomography, which reconstructs a single density operator, quantum process tomography seeks to identify the entire dynamical map governing a system’s evolution, requiring the estimation of a much higher-dimensional CPTP operator with $d^4-d^2$ free parameters.

In particular, a quantum process $\mathcal{E}:\rho\mapsto \mathcal{E}(\rho)$ admits an operator representaiton  in operator elements $E_i\in\mathbb{C}^{d\times d}$, e.g., Kraus decomposition as discussed in \sectionref{sec:Qchannels},
\begin{equation}
    \mathcal{E}(\rho)=\sum_i E_i\rho E_i^\dagger,\quad \sum_i E_i^{\dagger}E_i=I.\label{eq:operator_repr}
\end{equation}
For tomography, it is advantageous to expand the channel in a fixed orthonormal operator basis
$\widetilde{E}_i$, 
\[
E_i=\sum_{j}e_{ij}\widetilde{E}_j,\quad e_{ij}\in\mathbb{C}.
\]
Then, Eq.~\eqref{eq:operator_repr} can be expressed as follows 
\begin{equation}
    \mathcal{E}(\rho)=\sum_{ij} \widetilde{E}_i\rho \widetilde{E}_j^\dagger \chi_{ij},\quad \chi_{ij} = \sum_m e_{mi}e_{mj}^*.
\end{equation}
The coefficients $\chi_{ij}$ form the entries of the positive semi-definite Hermitian matrix, $\chi$, and the above decomposition is also known as $\chi$-matrix representation. Thus, $\varepsilon$ can be completely described by the $\chi$ matrix given a fixed operator basis $\widetilde{E}_i$, and the goal of the QPT is to reconstruct the entries of $\chi_i$. After enforcing complete positivity and trace preservation this results in estimating $d^4-d^2$ real parameters. As in QST, the reconstruction again reduces to solving a linear inverse problem subject to complete-positivity and trace-preservation constraints. 
However, QPT involves estimating a much higher-dimensional object, making it even more sensitive to noise and ill-conditioning. Consequently, QPT is formulated as a constrained inference problem, typically addressed using maximum-likelihood or Bayesian methods. As in QST, UQ plays a central role, providing confidence regions for the reconstructed parameters and propagating measurement uncertainty to derived operational metrics, such as fidelities, error rates, and channel distances used in benchmarking and control.

\textbf{Gate set tomography (GST)}, as formalized in~\cite{nielsen2021tomography} and building on the self-consistent framework introduced by~\cite{merkel2013}, provides a self-consistent estimator for the triple $(\rho, \{G_i\}, M)$,\linebreak[0] consisting of the prepared state, the gate operations, and the measurement effects, without assuming any of these components are known a~priori. 
In contrast to standard process tomography, where the prepared state $\rho$ and measurement $M$ are assumed to be fixed and known, causing SPAM errors to be indistinguishably mixed with gate dynamics, GST infers all components simultaneously. It does so by fitting a likelihood model to data from carefully designed gate sequences whose lengths amplify small coherent and stochastic errors. The resulting estimation problem is posed as a constrained statistical inference task, enforcing complete positivity and trace preservation across the entire gate set.
From a UQ perspective, GST is particularly powerful because it enables uncertainties in SPAM and gate parameters to be separated and consistently propagated. 

\textbf{Hamiltonian and Lindblad learning} is a structured form of quantum process characterization in which the dynamics are modeled at the level of a physical generator rather than an arbitrary quantum channel. The goal is to calibrate a parametrized master equation Eq.~\eqref{eq:master} in which coherent and dissipative contributions are explicitly represented and inferred from experimental data. Instead of estimating a general CPTP map, these approaches infer a low-dimensional set of physically interpretable parameters, such as frequencies, coupling constants, control amplitudes, and noise rates. A calibrated master equation offers a predictive model consistent across timescales and circuit depths, supporting accurate simulation, systematic calibration, and the design of error mitigation and correction strategies. 

In practice, the Hamiltonian is usually decomposed into, e.g.,
\begin{equation*}
    H(t) = H_{\mathrm{sys}}(t)+H_{\mathrm{ctr}}(t)+H_{\mathrm{err}}(t),
\end{equation*}
where $H_{\mathrm{sys}}(t)$ denotes the intrinsic system Hamiltonian, $H_{\mathrm{ctrl}}(t)$ the control Hamiltonian, and $H_{\mathrm{err}}(t)$ is the coherent error contributions. 
Then, the master equation Eq.~\eqref{eq:master} is written in a parametrized form, 
\begin{equation}\label{eq:master_param}
    \deriv{\rho(t)}{t} = \mathcal{L}_{\theta}(t)[\rho(t)]:=-i \left[H(t;\zeta),\rho(t)\right] + \sum_k\gamma_k\left( L_k \rho(t) L_k^\dag - \frac{1}{2} \left\{ L_k^\dag L_k, \rho(t)\right\}\right),
\end{equation}
with the set of unknown parameters $\theta=(\zeta,\gamma)$, corresponding, e.g., to system frequencies, couplings, control amplitudes, or noise rates. If the $L_k$'s are not also model parameters, they should be posited based on knowledge of the physics or constitute a complete basis. The goal is to learn these parameters from experimental data, yielding a calibrated dynamical model that enables predictive simulation of circuits, systematic calibration of controls, noise mitigation strategies, and the design of error correction schemes tailored to the actual error environment. 
Also, estimating $\theta$ from time-resolved measurements or circuit-level experiments yields a generator-level model from which entire families of quantum channels $\{\mathcal{E}_t\}$ can be predicted.

From a UQ perspective, Hamiltonian and Lindblad learning associate uncertainty directly with physically inter\-pretable parameters. Likelihood-based and Bayesian methods~\cite{granade2012robust} yield confidence regions or posterior distributions over~$\theta$, which can be propagated through the learned generator $\mathcal{L}_{\theta}$ to quantify uncertainty in predicted quantum channels and derived metrics.

\subsection{Error Management} \label{sec:error_management}

Error management in quantum computing encompasses strategies for understanding, reducing, and accounting for the impact of noise on algorithm performance. Near-term quantum devices operate in the presence of unavoidable imperfections—parametric uncertainties in gate calibrations, stochastic environmental fluctuations, and time-varying drift—that degrade computational outputs. Rather than relying solely on full quantum error correction, which remains prohibitively expensive for current hardware, practitioners employ a suite of UQ-informed techniques to manage these errors at multiple levels. Sensitivity analysis identifies which noise sources and parameter uncertainties most strongly influence algorithm outcomes, enabling targeted calibration and resource allocation. Error mitigation methods trade computational overhead for reduced bias in measured observables, with rigorous UQ ensuring that reported confidence intervals honestly reflect both the variance inflation and residual model errors introduced by mitigation. Advanced techniques such as importance sampling efficiently quantify rare failure events, while hierarchical noise decomposition methods separate overlapping stochastic processes acting on different timescales. Together, these approaches provide a statistically coherent framework for propagating device-level uncertainties into algorithm-level performance guarantees, bridging characterization data with end-to-end computational reliability.

\subsubsection{Sensitivity analysis}

A driving question of basic research needs in quantum computing is related to addressing errors sufficiently through the stack. Progress in this direction requires designing error resilience protocols that optimally address noise in near-term devices while maintaining reasonably low overhead costs as devices scale. A way to evaluate the ability of quantum algorithms to yield consistent and accurate results despite noise becomes essential. Here we discuss UQ methods that quantify and propagate parametric uncertainties through quantum circuits, enabling practitioners to understand which sources of variability most significantly affect algorithm output.

\textbf{Quantification of Gate Uncertainty via Sensitivity Analysis.} Quantum circuits depend on a collection of tunable parameters, typically gate rotation angles $\theta\in \mathbb{R}^d$, that specify the unitary evolution. The circuit implements a parameterized unitary map $U(\theta): \mathcal{H}_n \to \mathcal{H}_n$, transforming an initial state $\ket{\psi_0}$ to $U(\theta) \ket{\psi_0}$. Observable measurements yield expectation values
\begin{equation}
    \mu(\theta) = \tr\bigl(O \, U(\theta) \rho_0 U(\theta)^{\dagger}\bigr) = \langle O \rangle_\theta
\end{equation}
where $O$ is an observable (Hermitian operator), $\rho_0$ is the initial state density matrix, and $\tr(\cdot)$ denotes the trace. In practice, hardware imperfections introduce uncertainties in gate parameters, so $\theta$ is drawn from a distribution with specified means and covariances. 

\emph{Global sensitivity analysis} quantifies how input uncertainties in $\theta$ propagate to output uncertainty in $\mu(\theta)$. Variance-based indices partition the output variance into components attributable to each parameter or parameter set. The most common framework uses Sobol' indices~\cite{xu2022quantifying}, which decompose the variance as
\begin{equation}
    \Var_\theta(\mu) = \sum_{i=1}^d S_i \Var_i + \sum_{i<j} S_{ij} \Var_{ij} + \cdots
\end{equation}
where $S_i$ is the \emph{first-order} (main effect) sensitivity index quantifying the individual contribution of $\theta_i$ to output variance, and $S_{ij}$ captures interaction effects between parameters. These indices satisfy $\sum_S = 1$ (after accounting for all interactions), providing a normalized ranking of parameter importance.

For quantum circuits, computing these indices typically involves:
\begin{enumerate}
\item \textbf{Sampling}: Draw ensembles $\{\theta^{(k)}\}_{k=1}^N$ from the input distribution $p(\theta)$, with sample size $N$ chosen to achieve desired accuracy.

\item \textbf{Evaluation}: For each sample, execute (or simulate) the circuit and measure $\mu(\theta^{(k)})$, recording the output. In experiments, this involves repeated measurements (shots) since quantum measurement is inherently stochastic. Simulation tends to be prohibitively costly except for small circuits or circuits restricted to special classes, such as Clifford circuits.

\item \textbf{Aggregation}: Compute empirical variance and conditional expectations to estimate Sobol' indices. 
\end{enumerate}

An alternative local approach expands $\mu(\theta)$ around a nominal point $\theta_0$ via Taylor series,
\begin{equation}
    \mu(\theta_0 + \Delta\theta) \approx \mu(\theta_0) + \sum_i \frac{\partial \mu}{\partial \theta_i}\bigg|_{\theta_0} \Delta\theta_i + \frac{1}{2} \sum_{i,j} \frac{\partial^2 \mu}{\partial \theta_i \partial \theta_j}\bigg|_{\theta_0} \Delta\theta_i \Delta\theta_j + \cdots
\end{equation}
If parameter perturbations $\Delta\theta$ are small, linear (first-order) approximations often suffice. The gradient (or Jacobian) $\nabla_\theta \mu$ can be computed via parameter-shift rules~\cite{xu2022quantifying}, a quantum-native differentiation technique that avoids classical finite-difference errors.

For small Gaussian perturbations, the propagated output uncertainty is approximately
\begin{equation}
    \Var(\mu) \approx \nabla_\theta \mu^T \Sigma_\theta \nabla_\theta \mu,
\end{equation}
where $\Sigma_\theta = \Var(\theta)$ is the input covariance matrix. The \emph{relative importance} of parameter $\theta_i$ is then $\frac{|\frac{\partial \mu}{\partial \theta_i}| \sigma_{\theta_i}}{\sqrt{\Var(\mu)}}$, indicating which parameter perturbations dominate output scatter.

Bayesian statistical approaches, such as the maximum entropy method, can integrate these sensitivities with empirical observations from a finite set of quantum measurements. By comparing posterior distributions (updated via measurement data) with theoretical predictions, one quantifies model discrepancy and calibrates input uncertainty distributions.

\textbf{Quantification of Algorithmic Failure via Importance Sampling.} Beyond parameter sensitivity, one must also quantify the impact of stochastic noise processes on algorithm success. For instance, a variational algorithm may fail to find the ground state, or the output of a quantum state preparation circuit may have unacceptably large infidelity. These binary or threshold events have low probabilities in well-designed settings, making direct Monte Carlo sampling inefficient.

Let $X_k$ denote the outcome of measurement shot $k$, and $Y = \mathcal{A}(X_1, \ldots, X_N)$ be a functional of the measurement sequence (e.g., a quality metric or success indicator). The algorithm fails if $Y$ falls below a threshold $t$, so the failure rate is
\begin{equation}
    p_{\text{fail}} = \P(Y < t) = \mathbb{E}\bigl[1_{Y < t}\bigr].
\end{equation}
Standard sampling requires roughly $1/p_{\text{fail}}$ trials to observe a single failure event; for rare events ($p_{\text{fail}} \ll 1$), this becomes prohibitively expensive.

Dynamical subset sampling~\cite{heussen2024dynamical} adaptively concentrates resources on the sequences most likely to cause failure by reweighting samples according to an importance distribution $q(\mathbf{x})$ that shifts probability toward failure-inducing configurations. If $p(\mathbf{x})$ is the original (rare) distribution, the estimator
\begin{equation}
    \widehat{p}_{\text{fail}} = \frac{1}{M} \sum_{m=1}^M 1_{Y_m < t} \frac{p(\mathbf{x}_m)}{q(\mathbf{x}_m)},
\end{equation}
where samples $\mathbf{x}_m \sim q(\cdot)$, remains unbiased while achieving reduced variance compared to naive sampling. To maintain well-defined confidence intervals and statistical rigor, one must carefully track the likelihood ratios (importance weights) and account for their contribution to total variance. This is critical for reporting accurate credible intervals around the failure probability estimate.

\subsubsection{Error mitigation}

Error mitigation (EM) seeks to reduce bias in observables measured on near-term devices without the full overhead of quantum error correction. Core strategies include zero-noise extrapolation (ZNE)~\cite{temme2017_errormitigationshortdepth,endo2018_practicalqem}, probabilistic error cancellation (PEC)~\cite{mari2021_quantumprobablisticerrorcancellation}, symmetry verification and subspace expansion~\cite{cai2023_quantumerrormitigation}, virtual distillation~\cite{huggins2021vd}, and learning-assisted variants~\cite{liao2024_mlforpracticalqem,ferracin2024_efficientlyimprovingperformance,kim2025_errormitigationstabilizednoise,wang2025_nonmarcovianqem}. UQ is essential because each method introduces stochastic overhead and modeling assumptions that must be reflected in the reported error bars. 

Across these methods, we model the measured observable as $\mu(\lambda)=\tr\!\bigl(O\,\Lambda_{\lambda}(\rho)\bigr)$, where $\Lambda_{\lambda}$ is a noise channel of strength $\lambda$. Error mitigation constructs an estimator $\widehat{\mu}$ for the ideal value $\mu(0)$ by combining data from modified circuits, additional samples, or learned models. While these procedures reduce systematic bias, they introduce additional variance and potential model error. As a result, the reliability of $\widehat{\mu}$ depends not only on finite-shot noise, but also on uncertainty in noise characterization (e.g., calibration drift or learned parameters) and on the validity of assumptions such as perturbative expansions or Markovian noise. Quantifying how these sources propagate into confidence or credible intervals is therefore central to assessing the performance of any mitigation protocol.

\textbf{Zero-noise extrapolation (ZNE)} estimates the ideal expectation value $\mu(0)$ of an observable by combining empirical estimators $\widehat{\mu}(s_j\lambda)$ obtained at amplified noise levels $s_j\lambda$, where $\lambda$ denotes the physical noise strength and $s_j>1$ are scaling factors. Here, the ``hat'' indicates a finite-shot estimator of the expectation value,
\[
\widehat{\mu}(s_j\lambda)=\frac{1}{N_j}\sum_{i=1}^{N_j} X_i^{(j)},
\]
where $X_i^{(j)}$ are measurement outcomes sampled from the noisy circuit executed at effective noise $s_j\lambda$. Noise amplification can be implemented, for example, by \emph{gate folding}, in which unitary gates $U$ are replaced by sequences such as $U U^\dagger U$ that ideally act as $U$ but increase the accumulated noise~\cite{temme2017_errormitigationshortdepth}.

Assuming the expectation value admits an expansion
\[
\mu(\lambda)=\mu(0)+a_1\lambda+a_2\lambda^2+\cdots,
\]
Richardson (polynomial) extrapolation constructs coefficients $c_j$ satisfying the moment constraints
\[
\sum_{j=0}^m c_j s_j^k = \delta_k, \quad k=0,\dots,m,
\]
so that the first $m$ orders in $\lambda$ are canceled. This yields the unbiased estimator (up to higher-order terms)
\[
    \widehat{\mu}_{\mathrm{ZNE}} = \sum_{j=0}^m c_j\,\widehat{\mu}(s_j\lambda),
\]
for which the bias scales as $\mathcal{O}(\lambda^{m+1})$. Under independent sampling across noise levels, the variance propagates as
\[
\Var(\widehat{\mu}_{\mathrm{ZNE}})=\sum_{j=0}^m c_j^2\,\Var\bigl(\widehat{\mu}(s_j\lambda)\bigr)
\approx \sum_{j=0}^m \frac{c_j^2\,\sigma_j^2}{N_j},
\]
where $\sigma_j^2$ is the outcome variance at scale $s_j\lambda$. Thus, while higher-order extrapolation reduces bias, it amplifies statistical uncertainty through large coefficients $c_j$, leading to a bias--variance tradeoff in the choice of $(s_j,c_j)$~\cite{endo2018_practicalqem}. In practice, one may instead fit a parametric model (e.g., polynomial or exponential decay motivated by noise accumulation) to the data $\{(s_j,\widehat{\mu}(s_j\lambda))\}$, or use bootstrap resampling to construct confidence intervals that account for other types of noise, such as shot noise and/or device drift.

\textbf{Probabilistic error cancellation (PEC)} is suprisingly simple: combine many noisy experiments in a clever way so that, on average, the errors cancel out. More concretely, suppose a target channel $\Lambda$ admits a quasi-probability decomposition in an implementable gate set $\{\Gamma_k\}$,
\begin{equation*}
    \Lambda = \sum_{k} q_k\,\Gamma_k, \qquad \gamma := \|q\|_1 = \sum_k |q_k| \ge 1.
\end{equation*}
Executing randomized gates $\Gamma_{k_r}$ with probabilities $|q_k|/\gamma$ and outcome signs $\operatorname{sgn}(q_k)$ yields an unbiased estimator
\begin{equation*}
    \widehat{\mu}_{\mathrm{PEC}} = \frac{1}{N}\sum_{r=1}^N \gamma\,\operatorname{sgn}(q_{k_r})\,o_r, \qquad \EXP[\widehat{\mu}_{\mathrm{PEC}}]=\tr\bigl(O\,\Lambda(\rho)\bigr),
\end{equation*}
where $o_r$ is the measurement outcome on shot $r$. The variance scales as $\Var(\widehat{\mu}_{\mathrm{PEC}})=\gamma^2\Var(o_r)/N$, so $\gamma$ (the quasi-probability 1-norm) is a variance inflation factor often exponential in circuit depth. Importance-sampling bounds, stratified sampling, and control variates temper this growth~\cite{mari2021_quantumprobablisticerrorcancellation}.

\textbf{Symmetry verification and subspace expansion.} Suppose the ideal (noise-free) circuit prepares a state $\rho_0$ that lies entirely in a symmetry sector defined by a Hermitian operator $S$ with eigenvalues $\pm 1$ (e.g., a parity operator), so that $S\rho_0=\rho_0 S=\rho_0$. The projector onto the correct symmetry subspace is
\begin{equation*}
    \Pi = \frac{1}{2}(I+S), \qquad \Pi^2=\Pi.
\end{equation*}
On noisy hardware, the prepared state is $\rho=\Lambda_\lambda(\rho_0)$, which may have components outside this subspace. 

\emph{Symmetry verification} consists of measuring $S$ (or equivalently $\Pi$) and postselecting on the outcome $+1$. The resulting (normalized) postselected state is
\begin{equation*}
    \rho_{\mathrm{sym}} = \frac{\Pi \rho \Pi}{\tr(\Pi \rho)}.
\end{equation*}
Thus, for an observable $O$, the ideal expectation value is estimated by
\begin{equation*}
    \langle O \rangle_{\mathrm{sym}} 
    = \tr(O \rho_{\mathrm{sym}})
    = \frac{\tr(\Pi O \Pi \, \rho)}{\tr(\Pi \rho)}.
\end{equation*}
In an experiment with $N$ shots, this corresponds to keeping only those outcomes with $\Pi=1$, giving the estimator
\begin{equation*}
    \widehat{\mu}_{\mathrm{SV}} 
    = \frac{1}{N_{\Pi}} \sum_{r:\Pi=1} o_r,
    \qquad 
    N_{\Pi}\sim \mathrm{Binomial}(N,p_\Pi),
\end{equation*}
where $p_\Pi=\tr(\Pi \rho)$ is the \emph{pass probability}. Since only a fraction $p_\Pi$ of samples are retained, the variance scales as
\begin{equation*}
    \mathrm{Var}(\widehat{\mu}_{\mathrm{SV}}) \approx \frac{\sigma^2}{N p_\Pi},
\end{equation*}
so postselection increases variance by a factor $1/p_\Pi$, while reducing bias by removing symmetry-violating error components. This bias–variance trade-off is a general feature of quantum error mitigation~\cite{cai2023_quantumerrormitigation}.

\emph{Subspace expansion} generalizes this idea by replacing the single projector $\Pi$ with a small set of operators $\{G_i\}$ (for example, $G_0=I$ and $G_1=S$). One constructs an improved effective state
\begin{equation*}
    \rho_{\mathrm{sub}} 
    = \frac{\Gamma \rho \Gamma^\dagger}{\tr(\Gamma \rho \Gamma^\dagger)},
    \qquad 
    \Gamma = \sum_i w_i G_i,
\end{equation*}
where the coefficients $w_i$ are chosen (classically) to minimize the energy or another objective. The required quantities are the matrix elements
\begin{equation*}
    H_{ij} = \tr(G_i^\dagger O G_j \rho), 
    \qquad 
    S_{ij} = \tr(G_i^\dagger G_j \rho),
\end{equation*}
which define a small generalized eigenvalue problem whose solution yields the optimal weights $w_i$~\cite{cai2023_quantumerrormitigation}.

In this sense, symmetry verification is the special case where $\Gamma=\Pi$, while subspace expansion allows a richer linear combination of operators to partially correct errors that do not strictly violate the symmetry. Both methods reduce bias by projecting (exactly or approximately) back into a physically motivated subspace, at the cost of increased sampling overhead.

\textbf{Virtual distillation} uses multiple noisy copies to amplify the most dominant (or ``correct'') component of the state to suppress unwanted components, effectively filtering out incoherent noise. Suppose a noisy quantum circuit prepares a mixed state $\rho$ that is close to a desired pure state $\ket{\psi}$, for example
\[
\rho = F \ket{\psi}\!\bra{\psi} + (1-F)\sigma,
\]
where $F=\bra{\psi}\rho\ket{\psi}$ is the fidelity and $\sigma$ represents errors. Writing the spectral decomposition $\rho=\sum_i p_i \ket{\phi_i}\!\bra{\phi_i}$, one defines the $m$-copy ``distilled'' state
\[
\tilde{\rho}_m = \frac{\rho^m}{\tr(\rho^m)} 
= \frac{\sum_i p_i^m \ket{\phi_i}\!\bra{\phi_i}}{\sum_i p_i^m}.
\]
Since $p_i^m$ decays rapidly for smaller eigenvalues, the dominant eigenvector (ideally $\ket{\psi}$) is amplified. In fact, as $m$ increases, $\tilde{\rho}_m$ approaches the pure state corresponding to the largest eigenvalue exponentially fast~\cite{huggins2021vd}.

To estimate expectation values $\mu_m = \tr(O \tilde{\rho}_m)$, one does not explicitly prepare $\rho^m$. Instead, one uses $m$ copies of $\rho$ and the identity
\[
\tr(O \rho^m) = \tr\bigl(O^{(i)} S^{(m)} \rho^{\otimes m}\bigr),
\]
where $S^{(m)}$ is a cyclic permutation (generalized SWAP) operator acting on the $m$ copies. For the common case $m=2$, this reduces to measuring the SWAP operator $S$:
\begin{equation*}
    \widehat{\mu}_{\mathrm{VD}} 
    = \frac{\widehat{\tr}(O^{(1)} S\, \rho^{\otimes 2})}
           {\widehat{\tr}(S\, \rho^{\otimes 2})}
    = \frac{\widehat{\tr}(O\rho^2)}{\widehat{\tr}(\rho^2)}.
\end{equation*}
Intuitively, the SWAP test compares two copies of the state: if they are identical (high purity), the outcome is likely $+1$, while errors reduce this overlap. The statistical cost depends on the \emph{purity} $\tr(\rho^m)$. For $m=2$, the variance behaves approximately as
\[
\Var(\widehat{\mu}_{\mathrm{VD}}) 
\sim \frac{1}{R\,\tr(\rho^2)^2},
\]
where $R$ is the number of repetitions~\cite{huggins2021vd}. Thus, when the state is very mixed (small $\tr(\rho^2)$), many samples are required. The tradeoff here is that increasing $m$ improves bias (better approximation to a pure state) but increases sampling cost, since $\tr(\rho^m)$ can become exponentially small in system size or noise level.

\textbf{Learning-assisted mitigation and drift models.} A convenient way to formalize learning-based error mitigation is to view the quantum device as implementing a noisy map that depends on a set of (possibly hidden) parameters~$\theta$, such as calibration data or environmental variables.
Given a circuit~$C$ and observable~$O$, we write the measured expectation
\[
y = \langle O \rangle_{\mathrm{noisy}} = f(C,O;\theta) + \epsilon,
\]
where $\epsilon$ captures shot noise.
Then, the error mitigation strategy can reduce the noise based on these internal characteristics. It predicts an error mitigated value for the observable based on: the noisy value provided by the device (shot-noise and hardware-noise), relevant features of the circuit and observable underlying the algorithm, and the hidden parameters. It thus takes the form
\[
\langle O \rangle_{\mathrm{mitigated}}
= \widehat{g}(\langle O \rangle_{\mathrm{noisy}}, C, O; \theta)
\]
that predicts an error-mitigated value from features of the circuit \(C\), the observable, and the hidden, device dependent parameters \(\theta\). 
In practice, the former features may include gate counts, depth, and observables.
Machine learning (ML) methods with models ranging from kernel methods to random forests and neural networks have been proposed in~\cite{liao2024_mlforpracticalqem}, where the parameters of the noisy hardware have been replaced with internal hidden variables of the ML-model under consideration.
The hidden parameters are inferred by supervised training modalities, where training data is generated on a noise free simulator.

UQ arises naturally in this by treating the learned model as a probabilistic predictor.
In particular, a stochastic surrogate may not only predict the mitigated value from the noisy value, but provide information on the accuracy of that mitigated value (taking into account the aleatoric uncertainty caused by shot noise).

A key complication for any mitigation strategy depending on parameters \(\theta\) related to
noise processes is that the noise parameters are not fixed in time. Experiments show that device noise can fluctuate due to microscopic environmental effects (e.g., two-level systems), leading to time-dependent noise channels $\mathcal{E}_t$ and an observable drift~\cite{kim2025_errormitigationstabilizednoise}.
A simple mathematical model treats $\theta_t$ as a stochastic process, for instance a Gaussian Markov process,
\[
\theta_{t+1} = A \theta_t + \xi_t, \qquad \xi_t \sim \mathcal{N}(0,Q),
\]
so that the noisy expectation becomes
\[
y_t = f(C,O;\theta_t) + \epsilon_t.
\]
Learning-assisted mitigation must then average (or marginalize) over the posterior distribution of $\theta_t$:
\[
\widehat{\langle O \rangle} = \mathbb{E}_{\theta_t \mid \text{data}} \bigl[\, \widehat{g}(\langle O \rangle_{\mathrm{noisy}},C,O;\theta_t) \,\bigr],
\]
which inflates uncertainty when the posterior over $\theta_t$ is broad. More generally, when the environment exhibits memory, the noise is \emph{non-Markovian}, meaning the state evolution depends on its history. Mitigation error and sampling cost depend on correlations in the environment (e.g., bath correlation functions)~\cite{wang2025_nonmarcovianqem,kim2025_errormitigationstabilizednoise}. From a statistical perspective, this introduces temporal correlations in the data, so that effective sample size is reduced and confidence intervals widen.
In all these approaches, learning provides a map from \emph{observable noisy data} to \emph{ideal estimates}, while UQ quantifies how uncertainty propagates through (i) finite sampling, (ii) model approximation, and (iii) stochastic or correlated drift in the underlying noise. As drift becomes stronger or more non-Markovian, posterior uncertainty grows, and any reliable mitigation scheme must explicitly account for this through probabilistic modeling and marginalization.

\textbf{Cost–benefit and stopping rules.}
In practice, mitigation protocols are often preceded by hardware-level error-suppression or noise-shaping techniques such as dynamical decoupling and randomized compiling (e.g., Pauli twirling), which modify the effective noise channel $\Lambda_\lambda \mapsto \widetilde{\Lambda}_\lambda$. These methods do not themselves produce mitigated estimators, but can improve the validity of modeling assumptions and reduce baseline variance, thereby enhancing the performance of downstream QEM procedures. In this section, however, we had focused on estimator-level mitigation and treat such pre-processing as part of the experimental noise model.

Because QEM trades bias for variance, mitigation is worthwhile only when the estimated mean-squared error (MSE) satisfies
\begin{equation*}
    \mathrm{MSE}_{\mathrm{mit}} = \textrm{Bias}^2(\widehat{\mu}_{\mathrm{mit}}) + \Var(\widehat{\mu}_{\mathrm{mit}}) < \mathrm{MSE}_{\mathrm{raw}},
\end{equation*}
at the available shot budget. Pilot shots, bootstrap intervals, or Bayesian risk estimates can evaluate this inequality on-the-fly. Reporting mitigated observables together with an uncertainty budget (shot noise, fit/model error, calibration uncertainty, and variance blow-up factors such as $\gamma$ or $\sum_j c_j^2$) keeps QEM results reproducible and comparable across devices.

\begin{table}[htbp]
\centering
\small
\renewcommand{\arraystretch}{1.2}
\begin{tabular}{|p{3.2cm}|p{4.0cm}|p{4.4cm}|}
\hline
\textbf{Method} & \textbf{Typical variance scaling} & \textbf{UQ handles} \\ 
\hline
Zero-noise extrapolation & Mild--moderate increase from fit uncertainty; grows with extrapolation factor & Bootstrap/Bayesian fits to extrapolant; drift-aware weighting of noise-scaled points. \\ 
\hline
Probabilistic error cancellation & Variance multiplies by quasi-probability 1-norm squared ($\gamma^2$), often exponential in depth & Importance sampling bounds, stratified sampling, control variates; report $\gamma$ and effective sample size. \\ 
\hline
Symmetry verification / subspace checks & Mild shot loss from discarding symmetry-violating shots & Binomial/hypothesis tests on discard rate; propagate to confidence intervals on retained-sample estimators. \\ 
\hline
Learning-assisted mitigation (e.g., ML regressors, drift models) & Depends on model bias plus prediction variance & Cross-validation and posterior predictive checks; interval estimates that combine model and shot noise. \\ 
\hline
Virtual distillation & Variance grows with number of copies/shots and overlap estimation error & Delta-method or bootstrap error bars on overlap estimates; report depth/shot overhead explicitly. \\ 
\hline
\end{tabular}
\caption{Comparative view of common \emph{estimator-based} error-mitigation tools, emphasizing how variance inflation enters and which UQ techniques keep reported error bars honest.}
\label{tab:qem_variance}
\end{table}

These mitigation tools thus sit between characterization and algorithm execution: they consume calibration outputs (with their uncertainties) from benchmarking/tomography and produce downstream observables whose error bars remain faithful to both hardware noise and the variance overhead introduced by the mitigation itself, keeping the whole resource-characterization workflow statistically coherent.

\subsubsection{Hierarchical Discrete Fluctuation Auto-Segmentation (HDFA)}

HDFA is a hierarchical statistical framework for \textit{decomposing qubit noise into additive stochastic components acting on distinct timescales}. It begins with a measured time-series $f(t_r)$ and assumes that its fluctuations arise from a superposition of Random Telegraph Noise (RTN) processes and slowly varying drifts. RTNs are the simplest mathematical description for discrete, stochastic switching between two metastable physical states. A Hidden Markov Model (HMM) is a probabilistic model in which:
\begin{itemize}
\item there is an underlying hidden state sequence (a Markov chain), and
\item the observed data are generated from these hidden states through an emission distribution.
\end{itemize}

In HDFA, HMMs are used to model telegraph-like switching processes in qubit noise, where the hidden states represent discrete noise configurations (e.g., parity states) that cannot be directly observed. Standard HMMs cannot represent this structure without an intractable state space, so HDFA resolves this through a recursive, multiscale representation~\cite{agarwal2025fasttrackingdisentanglingqubitnoise}.

At the first level, $f(t_r)$ is modeled as
\[
f(t_r) = f^{(1)}_{c}(t_r) = s^{(1)}(t_r)+\frac{f^{(1)}_{\Delta}(t_r)}{2},
\]
where $s^{(1)}(t_r)\in\{-1,1\}$ is a hidden Markov process. HDFA tries to infer the most likely state trajectory $s^{(1)}(t_r)$ and update corresponding parameters.

To capture slower fluctuations, the algorithm removes the fastest component by defining
\[
f^{(2)}(t_r) = f^{(1)}_{c}(t_r),
\]
and repeats the same RTN--HMM procedure at the second level. This can be generalized in a recursive manner:
\[
f^{(n)}(t_r) = f^{(n)}_{c}(t_r) + s^{(n)}(t_r)+\frac{f^{(n)}_{\Delta}(t_r)}{2},
\]
\[
f^{(n+1)}(t_r) = f^{(n)}_{c}(t_r),
\]
until no further RTN structure is detected. This yields the multiscale decomposition
\[
f(t_r)
=
\sum_{n=1}^{N}
s^{(n)}(t_r)+\frac{f^{(n)}_{\Delta}}{2}
+
f^{(N+1)}_{c}(t_r),
\]
where each term corresponds to a distinct physical noise component with identifiable rates, amplitudes, and temporal structure. As a UQ method for quantum computing, HDFA provides a principled means to
\begin{enumerate}
\item separate overlapping stochastic processes,
\item estimate parameters governing each source of qubit noise, and
\item propagate these estimates into improved device models, calibrations, and error-mitigation strategies.
\end{enumerate}
This mathematically structured decomposition supports reproducible, interpretable uncertainty quantification in realistic, time-varying quantum environments.

\subsubsection{Langevin-based MCMC}

Stochastic Gradient Langevin Dynamics (SGLD) is a class of Markov Chain Monte Carlo (MCMC) methods for Bayesian inference in quantum noise char\-acter\-ization that iteratively samples from posterior distributions by combining gradient information with stochastic noise. Given measurement data $Y$ and noise model parameters $\theta$ (e.g., Pauli-Lindblad coefficients $\{\lambda_k\}$), SGLD requires an unbiased estimator of the gradient of the log posterior, $\hat{\nabla} \log \P(\theta|Y)$, where $\mathbb{E}[\hat{\nabla} \log \P(\theta|Y)] = \nabla \log \P(\theta|Y)$.

For quantum systems, constructing such estimators presents a fundamental challenge. The likelihood function $\P(Y|\theta)$---the probability of observing measurement outcomes $Y$ given noise parameters $\theta$---contains a normalizing constant (or partition function)
\begin{equation}
    Z(\theta) = \sum_{\text{all outcomes}} \P_{\text{unnorm}}(Y|\theta)
\end{equation}
that depends on $\theta$ and requires summing over exponentially many possible measurement outcomes. Computing $Z(\theta)$ exactly is computationally prohibitive, which makes both the likelihood and its gradient intractable. This situation is called a \textit{doubly-intractable} inference problem: we cannot evaluate $\P(Y|\theta)$ directly, nor can we compute $\nabla_\theta \log \P(Y|\theta)$ analytically.

Recent advances address this challenge through a two-step procedure~\cite{bennink2019}: First, sequential Monte Carlo (SMC) samplers generate approximate samples from noisy quantum circuit simulations to construct \textit{consistent} estimators of the gradient---estimators whose bias vanishes asymptotically but may be biased for finite sample sizes. Second, \textit{debiasing techniques} apply statistical corrections to remove the remaining bias, producing genuinely unbiased gradient estimators $\hat{\nabla} \log \P(\theta|Y)$ that satisfy the required expectation condition for SGLD.

Real-world quantum noise is often spatially and temporally correlated, and can be modeled by sparse Pauli-Lindblad error models involving Pauli operators weighted by non-negative coefficients. The sparsity structure (e.g., restricting to local or two-local terms based on qubit connectivity) makes such models tractable yet expressive. Applying SGLD to this inverse problem yields posterior distributions over the coefficients rather than point estimates, thereby enabling rigorous quantification of uncertainty in hardware error rates, including means, variances, and full covariance structure. This probabilistic characterization provides quantitative statements such as ``the depolarizing rate on the CNOT gate between qubits 1 and 2 is $0.005 \pm 0.001$,'' supports enhanced error mitigation strategies by reducing residual bias and optimizing sampling overhead, guides targeted hardware diagnostics by identifying which noise sources are most prominent or variable, and informs adaptive calibration protocols where control pulses are adjusted in real-time to fluctuating noise environments.


\section{Opportunities and Challenges}
\label{sec:opportunities}

This report has introduced quantum computing from the perspective of uncertainty quantification, demonstrating that the fundamental probabilistic nature of quantum measurement, the omnipresence of hardware noise, and the statistical character of every quantum output make UQ not merely a diagnostic afterthought but a core piece of quantum computation itself. We have shown how classical UQ tools---Bayesian inference, Monte Carlo sampling, sensitivity analysis, probabilistic error propagation, and inverse problem formulations---directly address core quantum challenges: characterizing device noise, calibrating error models, designing robust algorithms, and certifying performance with rigorous confidence intervals. By reframing quantum computing as an inherently statistical discipline, we aim to bridge the conceptual gap between mathematicians and quantum scientists, revealing that the language and methods of modern UQ offer a natural and powerful lens through which to understand, validate, and advance quantum technologies.

As quantum hardware matures from small-scale NISQ prototypes toward fault-tolerant architectures and as quantum algorithms transition from proof-of-concept demonstrations to production workloads integrated with classical high-performance computing, the role of UQ will only grow in importance. In what follows, we outline key opportunities and open challenges across the quantum technology stack where mathematically rigorous uncertainty quantification can deliver transformative impact---and where mathematicians, with their expertise in probabilistic modeling, numerical analysis, and inverse problems, are uniquely positioned to lead.

\subsection{Quantum Hardware Benchmarking and Calibration}

Current benchmarking protocols (randomized benchmarking, cycle benchmarking, cross-entropy benchmarking) provide aggregate error metrics, but translating these into predictive, device-level uncertainty models remains an open challenge. UQ offers a path forward by embedding hardware characterization within a Bayesian or frequentist inference framework that quantifies not just point estimates of error rates but their full posterior distributions, temporal correlations, and spatial heterogeneity across qubit arrays. Mathematicians can develop adaptive experimental designs that minimize calibration overhead while maximizing information gain, rigorous stopping rules that balance measurement cost against statistical precision, and hierarchical models that separate drift, crosstalk, and SPAM errors with provable uncertainty guarantees. As devices scale to hundreds or thousands of qubits, sparse sensing and tensor-network-based calibration surrogates---drawing on compressed sensing, randomized linear algebra, and high-dimensional statistics---will become essential, and mathematicians trained in these methods will be critical contributors.

\subsection{Quantum--Classical HPC Integration}

The convergence of quantum and classical high-performance computing represents one of the most urgent and promising frontiers for UQ research. Hybrid quantum--classical algorithms (VQE, QAOA, quantum-enhanced optimization) already iterate between quantum processors and classical parameter optimizers, but current workflows treat quantum outputs as deterministic function evaluations, ignoring shot noise, calibration uncertainty, and correlations across circuit instances. A UQ-aware HPC integration pipeline would propagate uncertainties end-to-end: quantifying how quantum sampling error flows into classical gradient estimates, how optimizer stochasticity compounds hardware noise, and how to allocate computational budgets optimally between quantum shots and classical surrogate refinements. Mathematicians can formulate rigorous multi-fidelity frameworks that adaptively balance expensive quantum queries against cheap classical surrogates, develop variance-reduction techniques (control variates, importance sampling, Rao--Blackwellization) tailored to quantum cost functions, and design distributed uncertainty-aware schedulers that co-optimize quantum circuit execution and classical post-processing across heterogeneous computing resources. As quantum co-processors become integrated into national-scale HPC facilities, the ability to certify that hybrid workflows produce statistically valid, reproducible results with quantified error bars will be indispensable---and mathematicians will architect these certification pipelines.

\subsection{Quantum Algorithm Development and Validation}

Quantum algorithm research has traditionally emphasized asymptotic complexity and worst-case performance, often neglecting statistical variability, finite-sample behavior, and the impact of realistic noise. UQ can transform algorithm development by making uncertainty analysis a consideration: quantifying the probability that a variational ansatz converges to the global optimum, bounding the variance inflation incurred by error-mitigation overheads, and certifying that reported algorithmic advantages remain statistically significant under device drift and calibration uncertainty. UQ is also intertwined with computational complexity itself: in Monte Carlo methods, phenomena such as the sign problem lead to exponentially growing variance and runtime, making statistical uncertainty—not just gate count—a complexity bottleneck~\cite{troyer2005complexity}. Similarly, in the NISQ setting, computational power is inherently defined through repeated sampling of noisy circuits and classical post-processing, so algorithmic complexity must account for noise and sampling cost rather than idealized unitary query counts~\cite{chen2023complexity}. Mathematicians can formulate concentration inequalities and tail bounds for quantum estimators, develop probabilistic certificates for approximate solutions (e.g., PAC-style learning guarantees for quantum machine learning), and design noise-aware algorithm variants whose performance degrades gracefully and predictably as error rates increase. By treating quantum algorithms as stochastic programs whose outputs are distributions rather than deterministic values, UQ provides a vocabulary for comparing algorithms, setting performance benchmarks, and guiding resource allocation---ensuring that the next generation of quantum algorithms is not only theoretically elegant but also statistically robust and practically deployable.

\subsection{Quantum Networking and Distributed Quantum Information}

Quantum networks introduce new layers of uncertainty: entanglement distribution over lossy channels, probabilistic heralding of Bell pairs, and asynchronous state synchronization across remote nodes. UQ methods tailored to networked settings---sequential decision-making under uncertainty, real-time filtering and state estimation, and multi-agent coordination with partial observability---can optimize entanglement routing, design adaptive measurement strategies that maximize fidelity given stochastic link quality, and provide online confidence intervals on distributed quantum state fidelities. Mathematicians with expertise in stochastic control, queueing theory, and sequential Monte Carlo can develop the mathematical foundations for certifying quantum network performance, enabling quantum internet protocols that guarantee end-to-end fidelity bounds with explicit confidence levels and gracefully degrade under transient failures or adversarial noise.

\subsection{Quantum Sensing and Metrology}

Quantum sensors promise unprecedented sensitivity for magnetic field imaging, gravimetry, and clock synchronization, but realizing these gains requires optimal experimental design under resource constraints and rigorous uncertainty quantification of reported measurements. UQ provides the statistical scaffolding to translate raw quantum sensor data into calibrated estimates with error bars: Bayesian inference for parameter extraction from noisy quantum signals, adaptive sensing strategies that dynamically allocate measurement shots to maximize information, and multi-sensor fusion algorithms that combine heterogeneous quantum and classical data streams with propagated uncertainties. Mathematicians can formulate Cramér--Rao bounds for quantum metrology protocols, design sequential decision rules that optimize sensing trajectories in real time, and develop robust estimators that remain valid under model mismatch and environmental fluctuations---ensuring that quantum sensors deliver not just raw sensitivity but statistically certified, reproducible, and actionable measurements.

\subsection{Domain Science Applications: Chemistry, Materials, and Beyond}

Quantum computing's long-term promise lies in simulating complex quantum systems---molecular electronic structure, strongly correlated materials, lattice field theories---that exceed classical capa\-bilities. But translating quantum outputs into scientifically meaningful predictions requires end-to-end uncertainty quan\-ti\-fi\-ca\-tion: propagating hardware noise through variational eigen\-solvers into molecular energy uncertainties, quantifying the statistical confidence of phase-diagram predictions, and validating quantum simulation results against experimental data within rigorous inverse problem frameworks. Mathematicians can bridge the gap between quantum algorithm developers and domain scientists by formulating hierarchical UQ pipelines that start from device-level error models, propagate through algorithm-specific noise channels, and produce domain-relevant observables (reaction barriers, critical temperatures, material properties) with full uncertainty budgets. This will enable quantum simulations to meet the evidentiary standards of experimental science---where every prediction comes with error bars and every claim of advantage is statistically falsifiable---and position quantum computing as a trusted partner in computational discovery rather than an exotic black box.

\subsection{The Mathematical Opportunity}

Across all these frontiers, the recurring theme is clear: quantum computing is not a purely physics or engineering challenge but a deeply mathematical one, and the tools of uncertainty quan\-ti\-fi\-ca\-tion---rooted in prob\-ability theory, statistical inference, numerical analysis, and optimi\-zation---are central to realizing quantum technology's potential. Mathematicians bring unique strengths to this endeavor: the ability to formulate problems rigorously, to prove guarantees rather than heuristics, to design algorithms with quantifiable convergence and error bounds, and to build modular, composable frameworks that remain valid as hardware and algorithms evolve. As quantum computing transitions from exploratory research to engineered systems integrated into scientific and industrial workflows, the demand for mathematically trained UQ practitioners will surge. Mathematicians who engage with quantum computing now---learning the domain, translating quantum challenges into inverse problems and stochastic optimization tasks, and contributing rigorous statistical methods---will not merely participate in this transformation but lead it, ensuring that the quantum future is not only powerful but also predictable, reproducible, and trustworthy.

\section*{Acknowledgments}
The authors would like to thank Michael Parks for being supportive of this effort, and Yan Wang for his careful reading and constructive comments.

\bibliographystyle{siamplain}
\bibliography{references}

\end{document}